\documentclass[a4paper,11pt]{article}
\usepackage{jinstpub} % for details on the use of the package, please see the JINST-author-manual
\usepackage{soul}

\title{\boldmath Evaluation of the Response to Electrons and Pions in the Scintillating Fiber and Lead Calorimeter for the Future Electron-Ion Collider}

\author[a]{H.~Klest,}
\author[a,1]{M.~Żurek\note{Corresponding author.},}
\author[c]{T. D.~Beattie,}
\author[a]{M.~Jadhav,}
\author[a]{S.~Joosten,}
\author[d]{M.~Kerr}
\author[a]{B.~Kim,}
\author[a]{M.~H.~Kim,}
\author[a]{J.~Metcalfe,}
\author[c]{Z.~Papandreou,}
\author[a,b]{J.~Richards,}
\author[c]{J.~Zarling}

\affiliation[a]{Physics Division, Argonne National Laboratory, \\ 
9700 S Cass Ave, Lemont, IL 60439, USA}

\affiliation[b]{Department of Physics, University of Connecticut, \\ 
196 Auditorium Road, Storrs, CT 06269, USA}

\affiliation[c]{Department of Physics, University of Regina,  \\
3737 Wascana Parkway, Regina, SK S4S 0A2, Canada}

\affiliation[d]{Department of Physics, Mount Allison University,  \\
67 York Street, Sackville, NB E4L 1E4, Canada}

\emailAdd{zurek@anl.gov}

\abstract{
The performance of the Baby Barrel Electromagnetic Calorimeter (Baby BCAL)—a small-scale lead-scintillating-fiber (Pb/ScFi) prototype of the GlueX Barrel Electromagnetic Calorimeter (BCAL)—was tested in a dedicated beam campaign at the Fermilab Test Beam Facility (FTBF). This study provides a benchmark for the Pb/ScFi component of the future Barrel Imaging Calorimeter (BIC) in the ePIC detector at the Electron-Ion Collider (EIC). The detector response to electrons and pions was studied at beam energies between 4 and 10\,GeV, extending previous GlueX tests~\cite{Beattie:2018xsk, Leverington:2008zz} to a higher energy regime.

The calibrated detector exhibits good linearity within uncertainties, and its electron energy resolution meets EIC requirements. The data further constrain the constant term in the energy resolution to below 1.9\%, improving upon previous constraints at lower energies. Simulations reproduce key features of the electron and pion data within the limitations of the collected dataset and the FTBF test environment. Electron–pion separation in the test beam setup was analyzed using multiple methods, incorporating varying degrees of beam-related effects. The inclusion of longitudinal shower profile information enhanced the separation performance, underscoring its relevance for the full-scale BIC in ePIC. These results provide essential benchmarks for the Pb/ScFi section of the future BIC, validating detector simulations and guiding optimization strategies for electron–pion discrimination.
}


\begin{document}
\maketitle
\flushbottom

\section{Introduction} 
\label{sec:intro}

The future Electron-Ion Collider (EIC) will probe the fundamental properties of matter by examining the internal structure of nucleons and nuclei with unprecedented precision. Currently being developed at Brookhaven National Laboratory, the EIC and its primary experiment, ePIC (the electron-Proton/Ion Collider experiment), aim to address some of the most profound questions in modern nuclear physics, including the origin of proton mass and spin, as well as the behavior of dense gluonic matter~\cite{WhitePaper, NASReport, YellowReport}. 

The goals of the EIC place stringent demands on the electromagnetic calorimetry in ePIC. In the central barrel region, the electromagnetic calorimeter must distinguish scattered electrons—critical for determining process kinematics—from background pions. It must also measure photon energy and position to identify signatures from deeply virtual Compton scattering and neutral pion decays~\cite{YellowReport}. The barrel electromagnetic calorimeter is required to reconstruct photons in the energy range of 100\,MeV to 10\,GeV, achieving an energy resolution better than $10\%/\sqrt{E} \oplus (2\text{--}3)\%$~\cite{EICRequirements}. Additionally, it must support electron identification and enable electron--pion separation for energies between 1\,GeV and 50\,GeV.

To meet the requirements in the barrel region, imaging calorimetry has been chosen for the ePIC detector. By integrating Pb/ScFi layers with AstroPix Monolithic Active Pixel Sensors (MAPS)~\cite{Brewer:2021mbe,Steinhebel:2022ips,SUDA2024169762}, the ePIC Barrel Imaging Calorimeter (BIC) will deliver precise three-dimensional imaging of particle showers. The Pb/ScFi component is inspired by the design of the GlueX Barrel Calorimeter~\cite{Beattie:2018xsk, Leverington:2008zz} (BCAL), featuring scintillating fibers aligned parallel to the beam axis and read out at both ends with silicon photomultipliers (SiPMs) to measure particle pseudorapidity, complementing the precise spatial information from the AstroPix layers. More information about the BIC concept can be found in Refs.~\cite{ATHENA:2022hxb, Apadula:2022ees}. Unlike two-dimensional calorimeters, this radially segmented design provides substantially more information about the particles traversing the detector. The inherently three-dimensional images produced by this hybrid imaging calorimeter enable separation of electrons from pions at the required level for the EIC, as demonstrated in detailed detector simulations utilizing the 3D cluster profiles generated by the calorimeter~\cite{ATHENA:2022hxb}.

This article presents results from a test beam campaign conducted in June 2024 at the Fermilab Test Beam Facility (FTBF)~\cite{ftbfBeamOverview}, as part of the T1224 Experiment. The focus is on the Pb/ScFi component of the BIC, using a small-scale prototype of the GlueX Barrel Calorimeter to benchmark its response to electrons and pions. The prototype is shallower (15.5 radiation lengths at normal incidence versus 17.1) and less segmented (four longitudinal readout segments versus twelve) than the planned BIC section, providing a simplified testbed for evaluating detector performance and reconstruction techniques.

The tests employed a mixed beam of electrons and hadrons, covering an energy range between 4\,GeV and 10\,GeV. These results were compared to the energy resolution requirements for ePIC and used to validate \textsc{GEANT4} detector simulations, which are important for understanding the achievable pion rejection using the $E/p$ method\footnote{The $E/p$ method identifies electrons by comparing the energy deposited in the Pb/ScFi calorimeter ($E$) with the momentum ($p$) measured by the tracking system, where electrons exhibit $E/p \approx 1$, while pions and other hadrons typically have $E/p < 1$.} and for assessing the calorimeter’s performance in photon and electron energy reconstruction. The validation was performed using realistic simulations of the prototype, implemented within the same software environment used for the ePIC detector~\cite{battaglieri_2024_14675920}. These are used to compare test beam results against expectations and to ensure the reliability of the simulated detector response used to guide the design and implementation of the BIC.

The GlueX BCAL was extensively tested at lower energy ranges, below about 2.5\,GeV. The GlueX collaboration reports a BCAL energy resolution of $5.2\%/\sqrt{E} \oplus 3.6\%$, integrated over typical angular distributions for $\pi^0$ and $\eta$ meson production in the 0.5--2.5\,GeV energy range. It has been noted that the response of the calorimeter averaged over its length for this energy range is not described well by the formula $\sigma/E = p_0/\sqrt{E} \oplus p_1$, and the fitted parameters $p_0$ and $p_1$ are highly correlated\footnote{The choice of omitting the $p_3/E$ noise term by the GlueX collaboration in the resolution fit is motivated by the fact that this term does not improve the fit. Studies in a similar energy range found no measurable contribution from the $p_3/E$ component, with fits yielding values for $c$ consistent with zero and producing nearly identical stochastic and constant terms, due to the low electronics noise and low-rate conditions reported in~\cite{Leverington:2008zz}}.
As the energy resolution was obtained by fitting only the low-energy data, it cannot fully constrain the constant term~\cite{Beattie:2018xsk}. Therefore, the main focus of the results presented here is on the electromagnetic response at higher energies, up to 10\,GeV, required for photon energy measurements at ePIC. Moreover, the response to pions was also tested to ensure the realism of the simulations that provide estimates of the expected pion--electron separation for the future Pb/ScFi section of the BIC.

\section{Test Beam Setup}
\label{sec:setup}
\subsection{Pb/ScFi Calorimeter Prototype}
\begin{figure}[htbp]
\centering
\includegraphics[trim={0 80 0 20},clip,width=\textwidth]{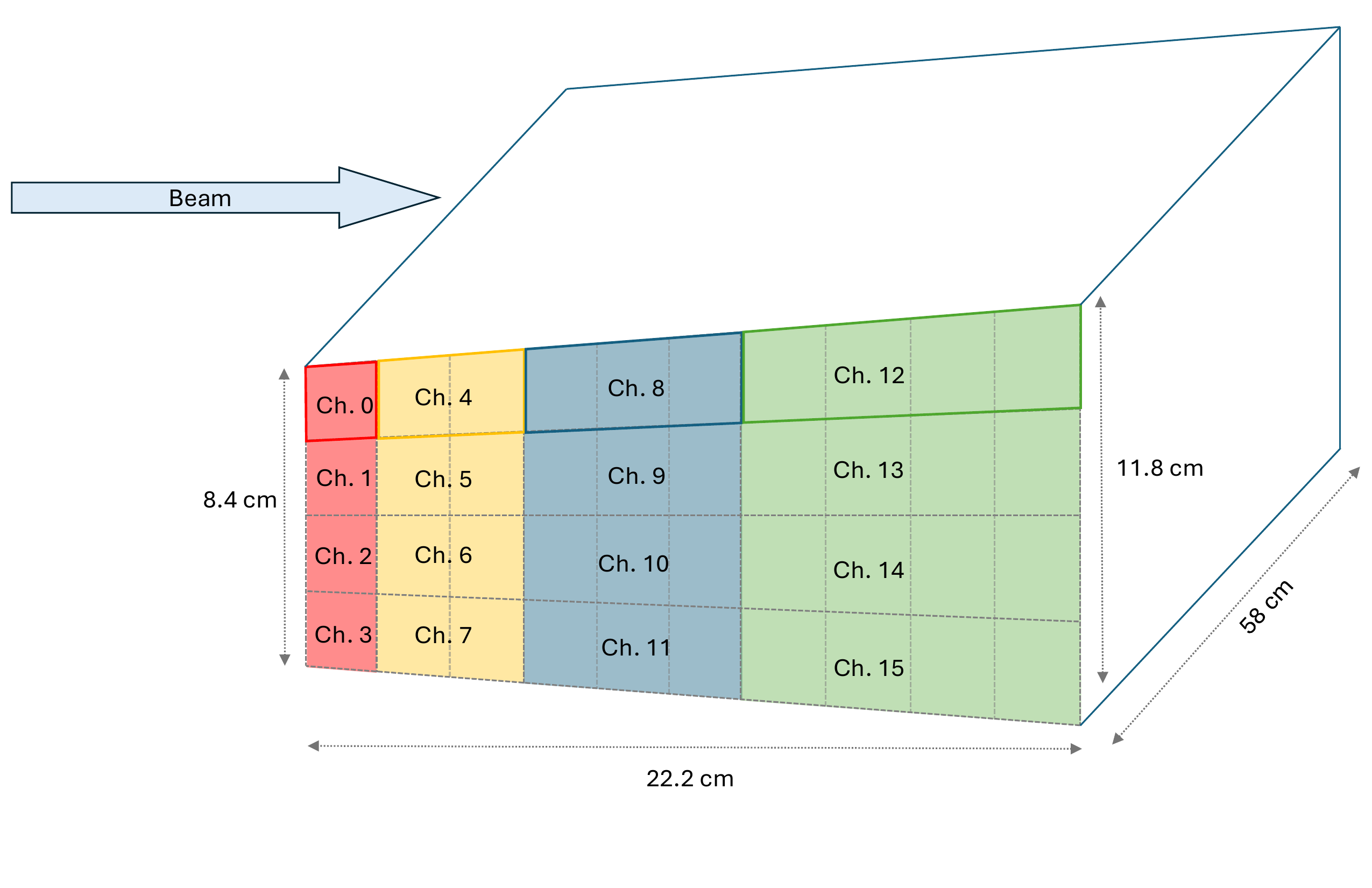}
\caption{The dimensions of the Pb/ScFi and the readout channel summing scheme of the Baby BCAL. The 40 SiPM arrays, concatenated into 16 readout channels, are outlined with dashed lines. Each column of ten SiPM arrays is divided into four readout channels; for example, column number 0, marked with bold solid lines, contains channels 0, 4, 8, and 12. The four rows of readout channels are marked in different colors: red, yellow, blue, and green, progressing from the most upstream to the most downstream position. The scintillating fibers run along the 58 cm long side of the Baby BCAL and are read out from both sides. In this schematic view, only one side (south) is visible; the channel numbering scheme is mirrored on the other (north) side of the prototype.}
\label{fig:ReadoutGeo}
\end{figure}

The prototype under test is a 58\,cm long segment of the GlueX BCAL~\cite{Beattie:2018xsk}, termed the \textit{Baby BCAL}. The Baby BCAL is in the shape of a trapezoidal prism with an active Pb/ScFi volume depth of 22.2\,cm, a major base of width 11.8\,cm, and a minor base of width 8.4\,cm (see Fig.~\ref{fig:ReadoutGeo}). In addition to the active volume, the design includes two aluminum plates: an inner plate on the minor base side that is 0.8\,cm thick, and an outer plate on the major base side that is 3.175\,cm thick. 

\begin{figure}[htbp]
\centering
\includegraphics[trim={80 110 95 100},clip,width=0.7\textwidth]{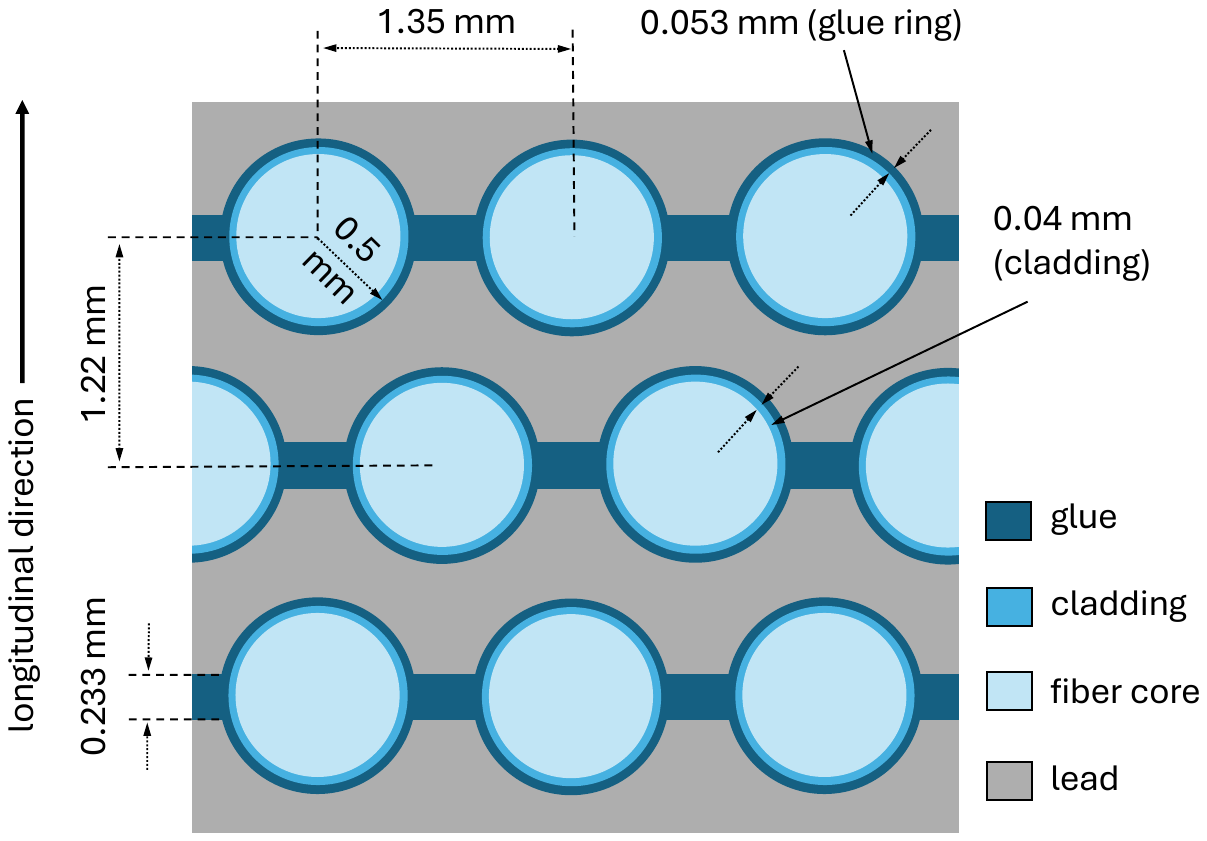}
\caption{Geometry of the Pb/ScFi matrix of Baby BCAL. The longitudinal direction is the direction of the shower development and the direction of the beam in Fig.~\ref{fig:ReadoutGeo}.}
\label{fig:ScifiMatrix}
\end{figure}

The active volume consists of 0.5\,mm thick corrugated lead sheets and epoxy surrounding 1\,mm diameter Kuraray SCSF-78MJ double-clad scintillating fibers~\cite{kuraray_psf}. The geometry of the Pb/ScFi matrix of the Baby BCAL is presented in Fig.~\ref{fig:ScifiMatrix}. The fibers are 58\,cm long and run along the length of the detector. The 184 sheets of lead were plastically deformed to create grooves to house the fibers. The sheets of lead and fiber are bonded together with BC-600 optical epoxy. The resulting Pb/ScFi matrix has an effective radiation length, $X_0$, of about 1.45\,cm, meaning the full prototype consists of approximately 15.5 radiation lengths at normal incidence. The prototype has a sampling fraction of approximately 9.5\%. The Molière radius is $R_M = 3.63\,\mathrm{cm}$, so the minor-base width of the prototype corresponds to about $2.3\,R_M$, and the major-base width to about $3.6\,R_M$~\cite{Beattie:2018xsk}.

The matrix of lead and scintillating fibers in the ePIC BIC is planned to follow exactly the geometry of the GlueX matrix described above and shown in Fig.~\ref{fig:ScifiMatrix}, making the Baby BCAL a good testbed to validate the simulation response for the BIC; the only difference is that single-clad fibers will be used instead of the double-clad fibers employed in GlueX. Figure~\ref{fig:setup} presents a picture of the Baby BCAL in the experimental setup commissioned during the FTBF tests presented in this article.

The light produced in the scintillating fibers travels to both ends (denoted north and south, respectively) of the calorimeter where it is guided to a $1.3 \times 1.3\,\text{cm}^2$ area of silicon photomultiplers (SiPM) by 40 acrylic light guides of 8 cm length. The photosensors are Hamamatsu S12045 SiPMs with 50 micron pixel size and a photon detection efficiency of around 28\% for 460 nm light at an overvoltage of 0.9 V~\cite{Soto:2014sca}. The output spectrum of the fibers peaks at 450 nm, which is well matched to the photosensitivity of silicon photomultipliers. The SiPM arrays themselves are a $4 \times 4$ array of $3 \times 3\,\text{mm}^2$ SiPMs. The SiPMs were temperature stabilized at 24\textdegree C via chilled water. To accommodate the trapezoidal shape, the shapes of the light guides are all unique. The interface between the light guide and the SiPM array is a 0.5 mm air gap. 

To reduce the overall number of readout channels in the GlueX BCAL, some of the SiPM-array signals are summed together into a single readout channel~\cite{Beattie:2018xsk}. This summing scheme is present also in the Baby BCAL readout. Each column of ten SiPM arrays is divided into four readout channels consisting of one, two, three, and four summed SiPM-array signals. The scheme employed to concatenate the 40 SiPM arrays into 16 readout channels is outlined in Fig.~\ref{fig:ReadoutGeo}. The most upstream row of four readout channels (marked in red in Fig.~\ref{fig:ReadoutGeo}) consist of a single SiPM array and the most downstream one (marked in green in Fig.~\ref{fig:ReadoutGeo}) consists of four summed SiPM arrays. 

The signals from the SiPMs pass through an on-board preamplifier and summing circuit before being sent via LEMO cable to Jefferson Lab 12-bit 250\,MHz sampling flash ADCs~\cite{jlabFADC250Manual} located in the FTBF electronics room. The flash ADCs operate in waveform mode, providing 112 samples of 4\,ns each for all 32 channels at every trigger. Typical waveforms are around 40 samples (160\,ns) long. No thresholds were applied to the SiPM signals with full waveforms read out for all channels at every trigger. The entire dataset presented in this paper was acquired using this zero-threshold configuration. The readout window is large enough that pedestals can be extracted and subtracted from the waveforms for each channel in each event. The dynamic range of the flash ADCs for measurement of the analog signal is 4096 ADC units. The pedestal-subtracted pulse height is used in the measurement of energy, as described in Sec.~\ref{sec:datacalib}.

\subsection{Fermilab Test Beam Facility}
\label{sec:FTBF}
% FTBF Beam
The calorimeter was stationed at the MTest beamline in the 6.2A enclosure. MTest can deliver beams of 120\,GeV protons, as well as mixed beams of electrons and hadrons produced from interactions of the proton beam on an upstream target. Furthermore, a dense absorber can be inserted into the beam to remove electrons, providing a beam of almost exclusively hadrons and muons.

The beam structure in MTest is such that a 4.2 second spill is delivered every 60 seconds. During the 4.2 second spill, every 11.2 microseconds, a 1.6 microsecond long batch is extracted from the main injector and sent down the MTest beamline. For intensities below 375,000 particles per spill, each batch arriving at MTest will on average contain less than one particle~\cite{ftbfBeamOverview}. The typical requested rate for this experiment was 50,000 particles per spill, resulting in a rate of around 12 kHz within the spill. The particles within the spill were typically spaced by more than 11.2 microseconds, which is long compared to the recovery time of the detector hardware. 

% FTBF Detectors    
The FTBF complex provides three scintillator counters located approximately 2, 5, and 15 meters upstream of the Baby BCAL setup in MTest 6.2A. The trigger for reading out the Baby BCAL consisted of hits in each of the two farthest counters, as the closest one was found to have poor efficiency.

In addition to the scintillators, FTBF provides two gas Cherenkov detectors with adjustable pressure settings. During data taking, both were operated in threshold mode with nitrogen gas to select electrons. The upstream detector uses a 24-meter-long gas radiator, with Cherenkov photons focused by a mirror onto a single photomultiplier tube (PMT). The downstream detector has a 15-meter gas radiator and directs photons onto one of two PMTs, depending on the Cherenkov angle. The inner PMT typically captures light from smaller angles, while the outer PMT detects photons emitted at larger angles from particles with $\beta \approx 1$.

\begin{figure}[ht]
 \centering
 \includegraphics[width=0.8\textwidth]{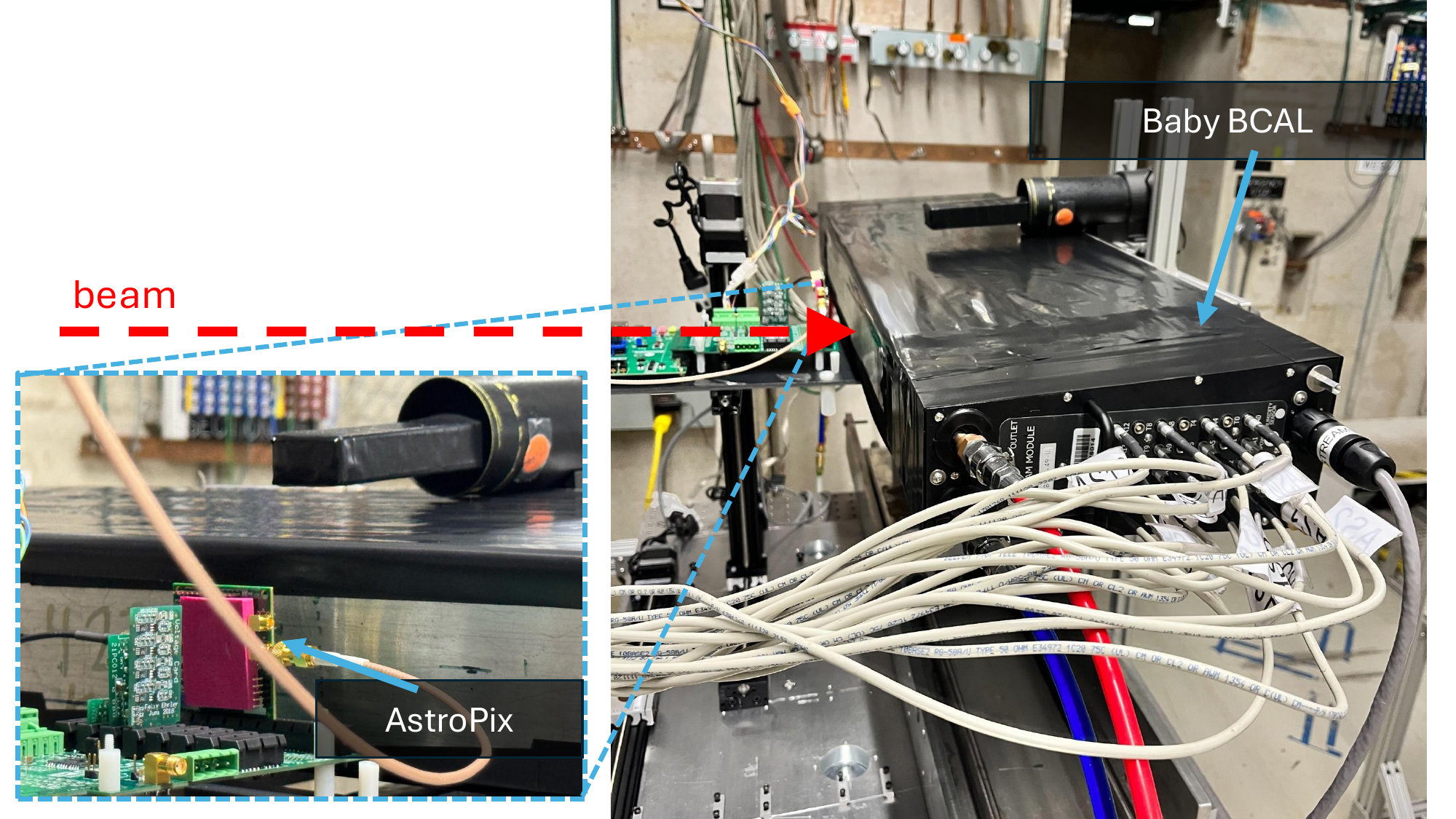}
\caption{The Baby BCAL setup at FTBF. The AstroPix sensor is aligned with the geometric center of the expected beamline and the center of the Baby BCAL. It is covered with a pink plastic covering which is larger than the size of the sensor itself. Note the orientation of the Baby BCAL; the set of 16 south-side white signal cables is visible in the right plot. The other side of the Baby BCAL has an analogous set of 16 north-side signal channels, enabling readout from both ends of the prototype.}
 \label{fig:setup}
\end{figure}

%FTBF Beam Profile
Due to the finite width of the calorimeter, showers that occur near the top or bottom (vertically) will experience significant energy leakage. The simulation of this effect relies on information about the profile of the beam at the calorimeter face. During the data collection period, the multiwire proportional chambers that typically provide the tracking for experiments at FTBF were unavailable. Therefore, a small AstroPix HV-MAPS silicon pixel sensor of $1.8 \times 1.8\,\text{cm}^2$ with a pixel pitch of $500 \times 500\,\mu\text{m}^2$~\cite{Brewer:2021mbe,Steinhebel:2022ips}, located approximately 2\,cm from the face of the calorimeter, was used to quantify the beam profile. The AstroPix chip was aligned with the FTBF laser positioning system to be centered both horizontally and vertically on the calorimeter face. The position of the AstroPix chip compared to the face of the calorimeter is shown in Fig.~\ref{fig:setup}.

\begin{figure}[htbp]
\centering
\includegraphics[width=.45\textwidth]{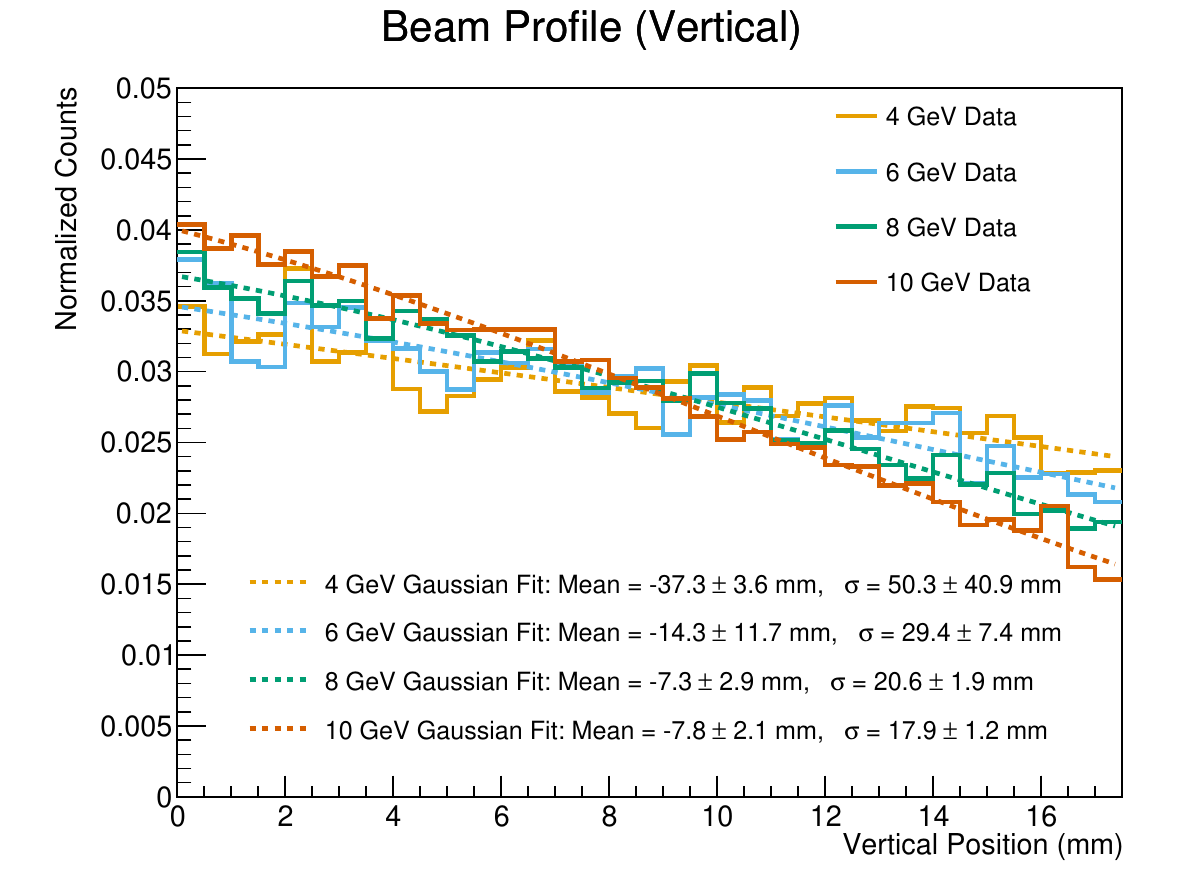}
\qquad
\includegraphics[width=.45\textwidth]{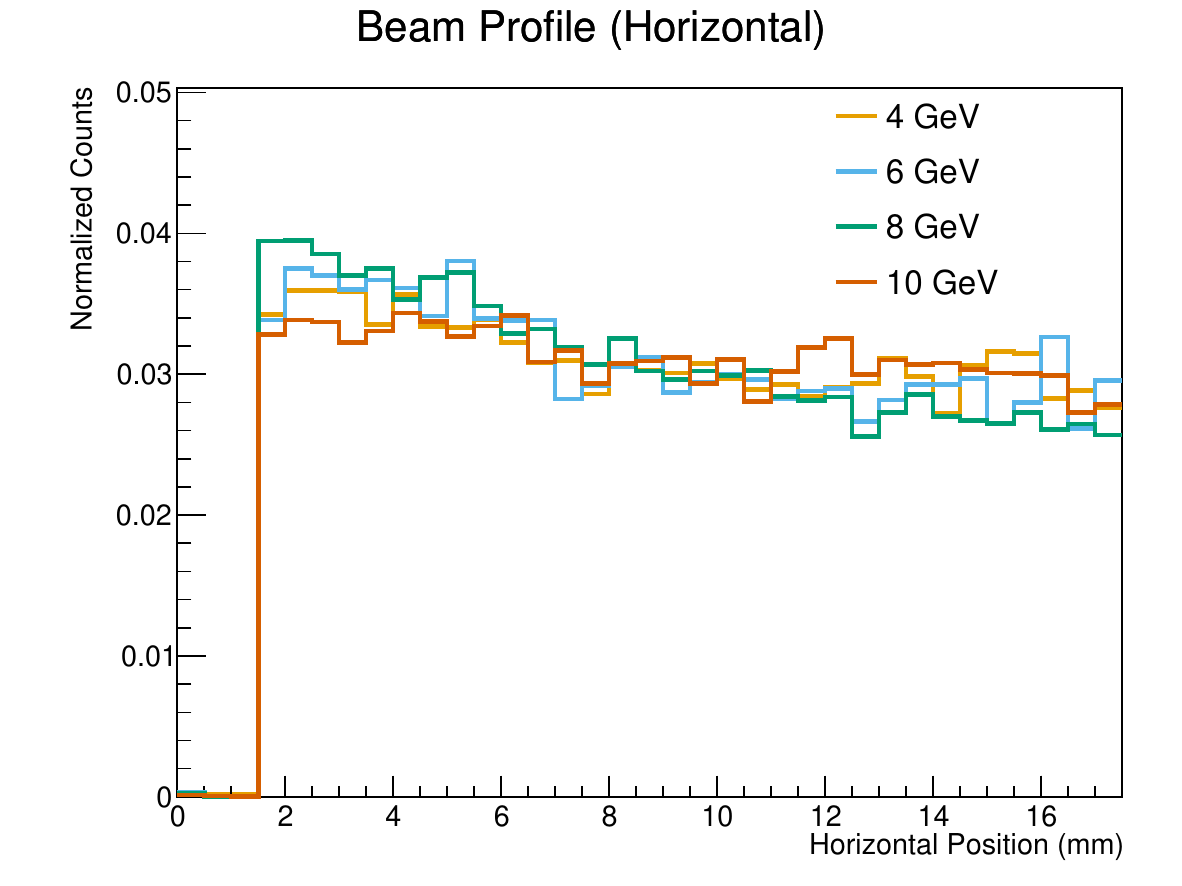}
\caption{Beam profiles as measured by the AstroPix sensor. Left: Beam profile in the vertical dimension. The distribution has been fit with a Gaussian function and the fit parameters are presented in the legend. Right: Beam profile in the horizontal dimension. The first three columns of pixels in the horizontal dimension are masked.}
\label{fig:APXBeamProfile}
\end{figure}

Despite the fact that the AstroPix is too small to measure the full beam profile, useful information can be gained by fitting the measured hit distributions for different beam energies. The measured hit distributions in the vertical and horizontal dimensions, together with fits to the vertical hit distributions, are shown in Fig.~\ref{fig:APXBeamProfile}. As can be seen in Fig.~\ref{fig:APXBeamProfile}, the beam profile in the vertical direction broadens at lower energy. Furthermore, the mean of the beam position is below the AstroPix. The first three pixel columns of the AstroPix v3 chip in the horizontal direction are equipped with PMOS instead of NMOS comparators, increasing the noise level. For the purposes of our test,
these columns were masked. No attempt was made to fit the horizontal profiles as they lack a clear energy dependence.

Since the FTBF mixed beam is produced by inelastic collisions on a fixed target upstream of MTest, there is additionally a non-negligible spread to the energy of the particles delivered, even at fixed steering magnet currents. This energy spread, which cannot be corrected for event-by-event, produces a non-negligible contribution to the measured energy resolution of our calorimeter. The resolution on the beam momentum $p$ was measured in the previous experimental runs at FTBF with a lead glass calorimeter. The standard deviations of the beam momentum spread ($\Delta p/p$) reported in the FTBF documentation~\cite{ftbfBeamMTest} are 2.7\% for the 4\,GeV beam and 2.3\% for the 8\,GeV. However, it should be noted that the beam is bent horizontally prior to arrival at MTest, resulting in a horizontal position dependence to the beam energy. Since the calorimeter under test is substantially longer in the horizontal dimension than the lead glass calorimeter, the energy spread of the beam is expected to be notably larger than cited. Furthermore, an energy offset of approximately 200\,MeV, constant with energy, was observed by the lead glass calorimeter measurements during the FTBF Experiment T1018 prior to our run period~\cite{tsai2024}. We therefore consider the effect of an energy offset of this magnitude throughout our results.

\section{Methods}
\label{sec:methods}
\subsection{Data Calibration}
\label{sec:datacalib}
Two different but complementary methods were used to convert the measured ADC units to energy for all 16 readout channels on both sides of the Baby BCAL. These methods are referred to as the ``MIP method'' and the ``electron method'', respectively. All the results presented in our analysis use the heights of the pulses measured by the flash ADCs as a proxy for the amount of energy deposited in a calorimeter cell.

\begin{figure}[htbp]
\centering
\includegraphics[width=.98\textwidth]{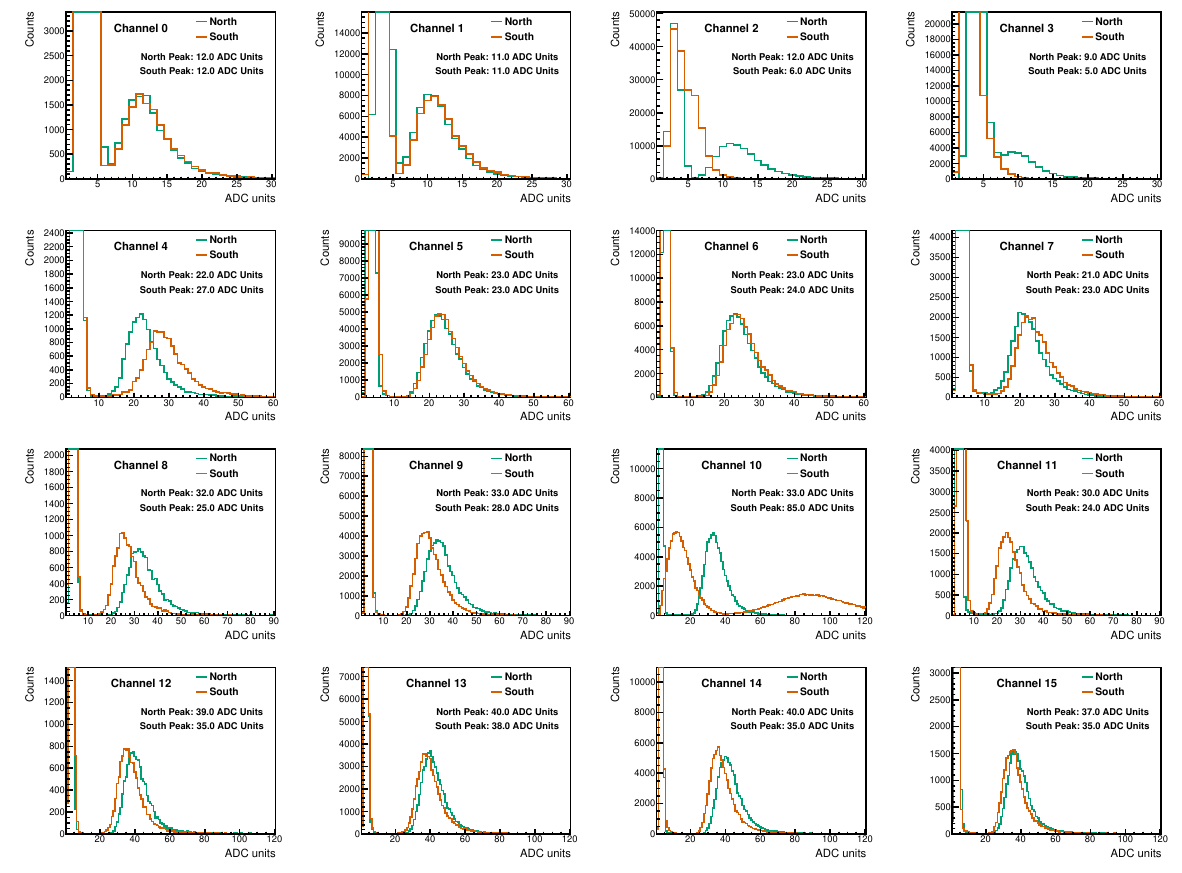}
\caption{Pulse amplitudes in ADC units for MIP-like events in the Baby BCAL. The peak positions are used to calibrate the detector. The low-ADC peak corresponds to the pedestal distribution. In channels S2 and S3, the MIP peak is not well resolved from the pedestal. In channel S10, the observed double-peak structure may result from a faulty amplification circuit.}
\label{fig:MIP_Peaks}
\end{figure}

The MIP calibration method utilizes the distinctive Landau peak in the deposited energy distributions provided by minimum ionizing particles. Since this peak corresponds to a well-defined energy deposit in the calorimeter material that can be determined analytically or from simulation, it can be used to align the relative responses of all the channels. The MIP calibration uses data collected with a 10\,GeV muon/pion beam, where a lead sheet was placed far upstream of the setup to absorb electrons, resulting in a pure sample of MIPs. Minimum ionizing particles are expected to leave straight tracks through the detector without showering (see e.g. Fig.~\ref{fig:EventDisplay4GeVPionShower}). As shown in Fig.~\ref{fig:MIP_Peaks}, a clear peak in the ADC distributions was observed in almost all channels of the detector for particles leaving only hits above threshold in a single column of channels. For two channels on the south side of the detector (S2 and S3), no MIP peak was observed. However, the impact of these channels on high-energy showering particles is minimal, since they reside at the very front of the calorimeter and correspond to only about 1.4 radiation lengths. The typical value of the MIP peak was approximately 10 ADC units per SiPM array, i.e. 10 units for the most upstream row and 40 units for the most downstream row. The estimated number of photoelectrons detected per SiPM array for MIPs hitting approximately the horizontal center of the Baby BCAL is 9-12. Once the MIP peak is found, it can be correlated with the true energy deposited by MIPs in the calorimeter determined by simulation. This provides a starting point to determine the conversion constants from ADC units to deposited energy. However, the sampling fraction generally differs between electron showers and MIPs~\cite{Livan:2017wnv}, meaning that the conversion between ADC units and the energy deposited \textit{by electrons} requires an additional multiplicative factor. This factor accounts for the ratio between the reconstructed electron energy (calibrated using MIPs) and the input beam energy. The derived value is approximately 14\%. In a full collider experiment, this same calibration procedure could also be accomplished using MIPs and the momentum information from electron tracks.

The electron method defines the channel calibration factors using the known incoming electron energy and a multi-differential minimization procedure. This procedure minimizes the difference between the calibrated energy reconstructed by each side of the detector and the nominal beam energy by optimizing the calibration constants. This method results in an overall better energy resolution compared to the MIP calibration method. However, since there are 16 calibration constants per side and only one beam energy, there is in general not a unique set of calibration constants. The optimization procedure used is differential evolution~\cite{storn1997differential}, with the MIP calibration constants used as an ansatz. To check the closure of this procedure, the electron beam data at 4, 6, and 8\,GeV beam energies were used to define the calibration constants, which could then be tested on the 10\,GeV data. The reconstructed peak of the 10\,GeV data was located at 9.9\,GeV and the resolution was slightly better than the MIP method, suggesting that this method produces calibration constants which are generalizable and not fine-tuned to a specific dataset. The final electron calibration constants were determined using the electron datasets at all four energies. As expected, compared to the MIP calibration method, the electron method produces reconstructed electron energy distributions with slightly better energy resolution (see e.g. Sec.~\ref{EEnergyReco}) and the set of calibration factors are generally similar. Channels which contribute little to the overall energy are poorly constrained by this method, however, these channels have essentially no impact on the results presented here. 

The studies on the linearity of the detector response compare both calibration methods and are presented in Sec.~\ref{sec:linearity}. For the measurements of electron and pion energies presented in this paper, we use the calibration constants extracted with the electron method. 

We evaluated the potential effect of signal attenuation in the scintillating fibers on our calibration and the extracted energy resolution. Attenuation reduces the light yield measured by the SiPMs as a function of the distance between the scintillation event and the readout. This effect is approximately exponential with distance, making it potentially significant for large distances. In our measurement, we do not perform an event-by-event horizontal-position-dependent energy calibration, as the exact horizontal position of each beam particle entering the Baby BCAL is not available. However, since the nominal beam position remained stable in the horizontal direction throughout the run period, the average attenuation is absorbed into the calibration constants determined separately for the north and south readout channels.

To estimate the potential impact of the poorly known horizontal beam spread on our results, we use the known dependence of the light yield on distance for the Baby BCAL fibers, as measured for GlueX \cite{Beattie:2018xsk}. In the most extreme scenario—signals originating at 0 cm from the north side and 58 cm from the south side—the signal attenuation factor is approximately 0.78. For signals originating at the center of the Baby BCAL, this attenuation factor is about 0.87. The total energy is reconstructed as the geometric mean of the north and south measured energies, $\sqrt{E_N \cdot E_S}$. Consequently, the difference in reconstructed energy for the extreme case of a signal produced at one end of the Baby BCAL compared to one produced at the center is only $\sqrt{1 \cdot 0.78} - \sqrt{0.87 \cdot 0.87} \approx 0.008$, or 0.8\%. It is important to note that the actual beam spread is expected to be much smaller, on the order of a few centimeters, compared to the extreme case of the full Baby BCAL length. Thus, we neglect the effects of horizontal beam spread in the subsequent analysis. For longer detectors, where attenuation effects are more significant, a shower-position-dependent attenuation correction would be necessary to optimize detector resolution.

\subsection{Particle Identification}
\label{sec:PID}
The primary datasets used in this analysis are mixed beams of electrons and pions at 4, 6, 8, and 10\,GeV. Using information from the FTBF Cherenkov detectors, we assigned a particle type to every event we registered in our data stream. The composition of the beam varies with energy, ranging from approximately 70\% electrons and 30\% pions at 4\,GeV to approximately 50\% electrons and 50\% pions at 10\,GeV when the negatively charged beam is selected~\cite{ftbfBeamComposition}. The 4\,GeV dataset is the largest and consists of 1.4 million triggers. The 8\,GeV dataset is smallest and consists of 320,000 triggers. To disentangle pions from electrons, the upstream and downstream Cherenkov detectors were used. They were operated at pressures of 2 and 3 psi of nitrogen, respectively. The two Cherenkov detectors can be checked against one another to determine their respective efficiencies. The upstream Cherenkov detector fired in 99.88\% of cases where the outer PMT of the downstream Cherenkov detector fired, and the outer PMT of the downstream Cherenkov detector fired in 99.91\% of cases where the upstream Cherenkov detector fired. The selection of pions was performed only with the requirement that neither the upstream Cherenkov detector or the outer PMT of the downstream Cherenkov detector were fired. This means that the pion sample contains an admixture of muons, kaons, and protons. However, since muons will typically not deposit large amounts of energy, and the fractions of kaons and antiprotons in the beam are known to be on the order of a few percent for negatively charged beams~\cite{ftbfBeamComposition}, particles which do not fire either Cherenkov detector but still leave a shower-like signal in the calorimeter are predominantly pions. 

\subsection{Channel-Level Analysis Cuts}
\label{sec:channel-level-analysis-cuts}

One channel of the Baby BCAL, channel 10 on the south side, exhibited non-linear behavior and rapid baseline fluctuations, likely due to damage in its amplifying circuit. This channel is located near the shower maximum for the studied energy range, making it crucial for shower energy reconstruction. Since the calorimeter has readout on both sides, problematic channels can be partially mitigated by relying on a single-side readout. In all subsequent results, the energy in channel 10 is taken as $E_{\text{N10}}$ rather than the geometric mean of both sides, $\sqrt{E_{\text{N10}}E_{\text{S10}}}$. Similarly, channels 2 and 3 are set to the north-side value due to calibration challenges with S2 and S3. However, since these channels receive only a small fraction of the shower energy, their impact is expected to be minimal. Example event displays of calibrated energy in Baby BCAL channels for identified electrons and pions, illustrating different energy deposit topologies, are presented in Appendix~A.

Various cuts are placed on events to ensure the event sample used to extract the energy resolution is free of confounding factors, such as transverse leakage and spurious signals. Events which have north and south channels that differ in the amount of reconstructed energy by more than 500 MeV are discarded. Additionally, showers should be approximately centered vertically on the calorimeter to ensure reasonable containment of energy. This is achieved by placing a cut which removes events where the highest measured energy deposit resides in the upper or lowermost columns of channels, namely channels 0, 4, 8, 12, 3, 7, 11, and 15. To reject showers occurring upstream of the calorimeter or in badly calibrated channels, we additionally reject events where the highest measured energy deposit is in the first row of channels, 0, 1, 2, and 3. These cuts together are henceforth denoted as the ``U'' cut, since they remove events with their highest energy deposits in a ``U'' shaped region.  

\subsection{Simulations}

\begin{figure}[htbp]
    \centering
    \includegraphics[width=0.6\textwidth]{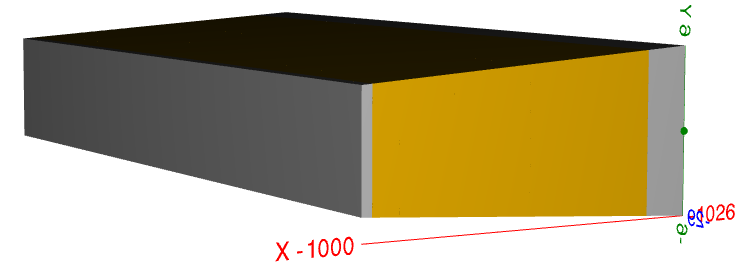}
    \caption{Geometry of Baby BCAL implemented in simulations. The gray volumes indicate the inner and outer aluminum plates, while the orange volume indicates the Pb/ScFi matrix, where fiber volumes with double cladding are placed within the epoxy and lead volume.}
    \label{fig:renderedGeo}
\end{figure}

The Baby BCAL responses are simulated in the official ePIC software environment~\cite{battaglieri_2024_14675920} using DD4hep to describe the detector geometry~\cite{frank_markus_2018_1464634} in tandem with GEANT4~\cite{GEANT4:2002zbu} version 11.0. The trapezoidal-shaped prototype geometry is implemented according to the dimensions presented in Fig.~\ref{fig:ReadoutGeo}, between the inner and outer aluminum plates as shown in Fig.~\ref{fig:renderedGeo}. The Pb/ScFi matrix is implemented according to the geometry presented in Fig.~\ref{fig:ScifiMatrix}, where fiber volumes with double cladding are placed within the epoxy and lead volume. The chemical compositions for the scintillating fibers, glue, and lead follow the GlueX BCAL reference~\cite{GlueX_fibers}. 

To ensure the simulation results closely match real experimental data, we employ a realistic digitization model that converts the energy deposits from the GEANT4 simulation into simulated responses of the readout electronics. We then apply reconstruction routines similar to those used for measured data, resulting in realistic simulated datasets analogous to our test beam data. We implemented an effective model for the light response, including light attenuation in the fibers, photoelectron statistics, light guide efficiency, and SiPM thresholds, based on previous beam and bench measurements with the Baby BCAL and optical simulations of the light guide response. We extracted the number of photoelectrons per unit of particle energy from measurements with a positron beam up to 6\,GeV in Hall D in 2023, similar to reference~\cite{Leverington:2008zz}. The short and long attenuation length components are taken from GlueX BCAL measurements~\cite{BAULIN201348, Beattie:2018xsk} and confirmed in recent bench measurements for the GlueX BCAL fibers. For each simulated event, we sum the deposited energy in the fibers corresponding to one SiPM array readout. We calculate the number of photoelectrons per GeV arriving at each of the 40 SiPM arrays on either end of the Baby BCAL from the known measured attenuation length parameterization, the distance of the shower center from the ends, and the attenuation-corrected number of photoelectrons per GeV measured in Hall D. We incorporate the relative differences in light guide efficiencies between individual channels. After smearing the number of photoelectrons per GeV according to Poisson statistics, we added electronic noise based on the typical pedestal observed in the data, and then converted the values into energy units. We consistently applied energy thresholds and cuts following the same approach outlined in Sec.~\ref{sec:channel-level-analysis-cuts}.

To account for the saturation of light produced by densely ionizing particles in our simulations, we adopt a Birks' constant of 0.132\,mm/MeV from the literature~\cite{POSCHL2021164865} for the scintillating fibers used in the Baby BCAL. This value is used in our GEANT4 simulation to convert deposited energy into scintillation signals. The FTFP\_BERT GEANT4 physics list is used in the simulations for the main results. The pion responses has been also simulated for comparison with the physics lists QGSP\_BERT, FTFP\_BERT, and their high precision versions (QGSP\_BERT\_HP, FTFP\_BERT\_HP). 

We account for the impact of the material in front of our setup, primarily consisting of FTBF Scintillator Counters and a component of the EMPHATIC experiment~\cite{EMPHATIC:2019xmc}, by including them in our GEANT4 simulations. The three FTBF Scintillator Counters (MT6SC1, MT6SC2, and MT6SC3) are implemented based on the parameters in~\cite{ftbfSC} and positioned as described in Sec.~\ref{sec:FTBF}. Additionally, a component of the EMPHATIC experiment located immediately downstream of MT6SC1 is modeled as an additional 2 cm of plastic scintillator material.

Table~\ref{tab:beam_conditions} presents the beam parameters used in the simulation, including two beam momentum spreads, $\Delta p/p$. The ``Cited'' values correspond to previous measurements by FTBF~\cite{ftbfBeamMTest} using the lead-glass calorimeter described in Sec.~\ref{sec:FTBF}. Values without an asterisk indicate direct measurements previously taken at the specified beam momentum. Values with an asterisk result from linear extrapolation between these measurements, since FTBF provides no direct data at these beam momenta. The ``Adjusted to the data'' values reflect the expected $\Delta p/p$ that best matches simulated Baby BCAL responses to measured electron energy resolutions. The $\sigma$ and mean of the beam spatial profile at the calorimeter surface are generated according to the beam profiles measured with the AstroPix sensor, shown in Fig.~\ref{fig:APXBeamProfile}. We assume the horizontal beam profile spread matches the vertical spread. See Sec.~\ref{sec:FTBF} for a discussion of the beam momentum spread and Sec.~\ref{sec:datacalib} for an estimation of the potential impact of uncertainties in the horizontal beam spread on energy reconstruction.

\begin{table}[h]
    \centering
    \caption{Beam parameters used in simulations. Two beam momentum spreads, $\Delta p/p$, are presented. Values labeled ``Cited'' refer to those provided by FTBF based on previous runs~\cite{ftbfBeamMTest}. Values without an asterisk were measured in the past for the given beam momentum, while those marked with an asterisk were obtained via linear extrapolation between measured points. The values labeled ``Adjusted to the data'' correspond to our inferred $\Delta p/p$ values required to match the simulated Baby BCAL energy resolution to the measured one.}
    \begin{tabular}{l c c}
        \hline\hline
        Beam momentum $p$ (GeV/$c$) & \multicolumn{2}{c}{Beam $\Delta p/p$ (\%)}  \\
        \hline
        & Cited & Adjusted to the data \\
        \hline
        4  & 2.7 & 3.3  \\
        6  & $2.5^{*}$ & 2.9 \\
        8  & 2.3 & 2.6   \\
        10 & $2.1^{*}$ & 2.4   \\
        \hline\hline
    \end{tabular}
    \label{tab:beam_conditions}
\end{table}

\section{Results}
\label{Sec:Results}
\subsection{Electron Energy Response}
\label{EEnergyReco}
\subsubsection{Energy Resolution}
\label{sec:energyres}
\begin{figure}[htbp]
\centering
\includegraphics[trim={3 0 50 30},clip,width=.47\textwidth]{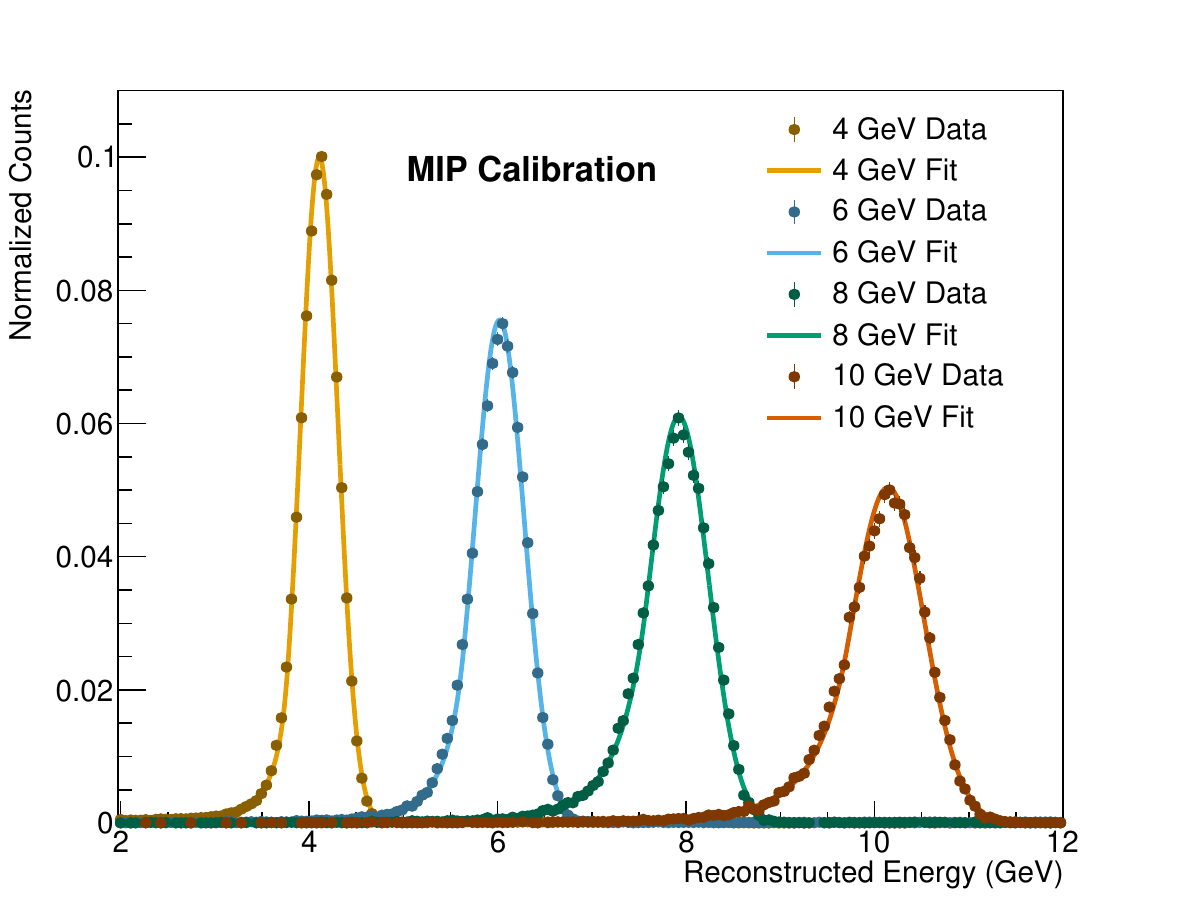}
\qquad
\includegraphics[trim={3 0 50 30},clip,width=.47\textwidth]{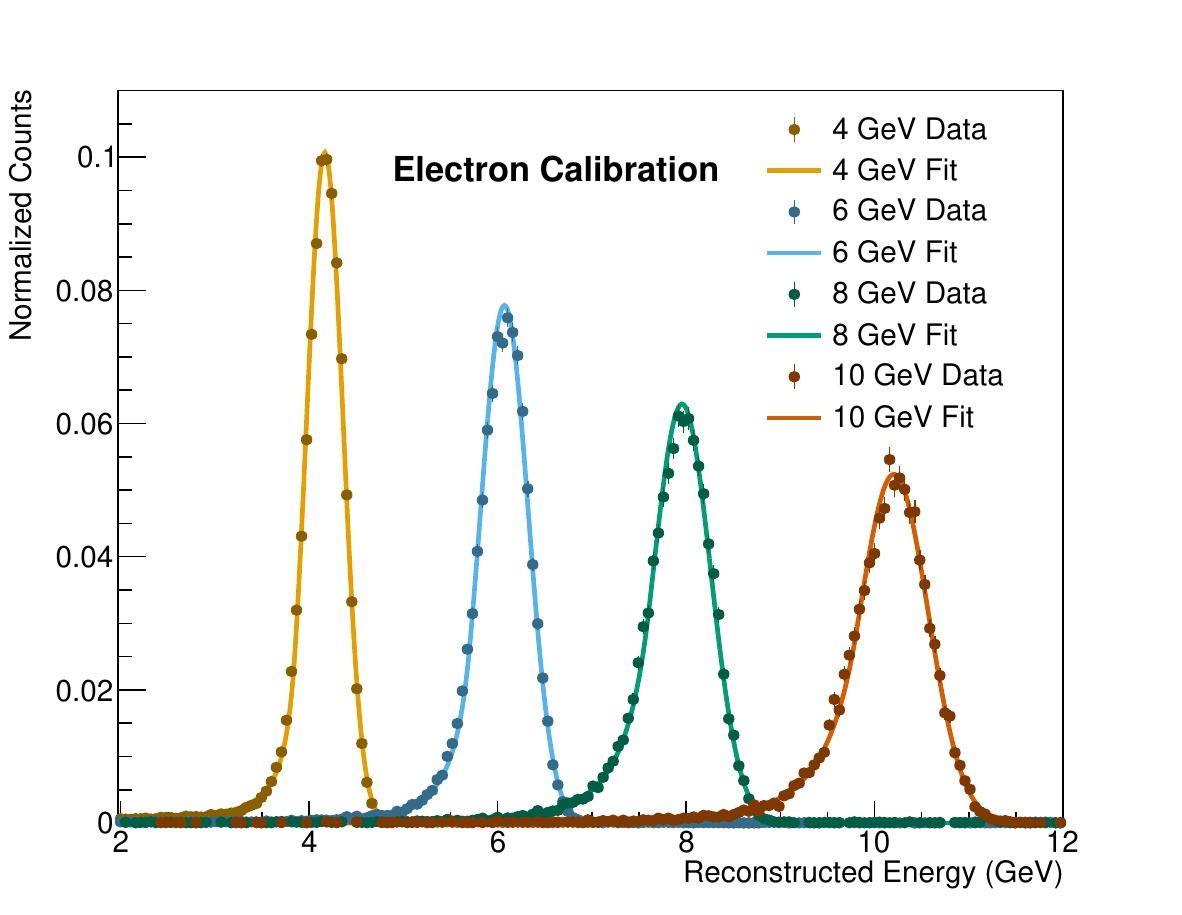}
\caption{Normalized distributions of measured electron energies at 4, 6, 8, and 10\,GeV in the Baby BCAL with Crystal Ball fits. Left: reconstructed energy distributions using the MIP calibration method. Right: same using the electron calibration method. See text for details on the calibration procedures.}
\label{fig:RecoEnergies}
\end{figure}

We reconstruct energies as the geometric mean of the energies reconstructed on the north and south sides, i.e., $\sqrt{E_N \cdot E_S}$. Figure~\ref{fig:RecoEnergies} presents the reconstructed electron energies for Cherenkov-selected electron events using two calibration methods, the MIP method on the left and the electron method on the right. In order to characterize and compare the energy performance between both methods, we fit with the measured energy deposits with a Crystal Ball function, extracting the peak position and the $\sigma$ of the Gaussian core (see Table~\ref{tab:fitparams}). The $\sigma$ value of this fit can be interpreted as the energy resolution $\sigma_E$.

\begin{table}[htbp]
\centering
\caption{Parameters of the Crystal Ball fits to the data calibrated using the MIP and electron methods, as shown in Fig.~\ref{fig:RecoEnergies}. The peak position and $\sigma$ of the Gaussian core are shown.}
\begin{tabular}{ccccc}
\hline\hline
Requested Beam Energy (\text{GeV}) & $\text{Peak}_{\hspace{0.2em}\text{MIP}}\ (\text{GeV})$ & $\sigma_{\hspace{0.2em}\text{MIP}}\ (\text{GeV})$ & $\text{Peak}_{\hspace{0.2em}\text{Ele.}}\ (\text{GeV})$ & $\sigma_{\hspace{0.2em}\text{Ele.}}\ (\text{GeV})$ \\
\hline
4.0 & 4.11 & 0.20 & 4.16 & 0.19 \\
6.0 & 6.02 & 0.26 & 6.07 & 0.25 \\
8.0 & 7.92 & 0.31 & 7.96 & 0.30 \\
10.0 & 10.14 & 0.38 & 10.21 & 0.35 \\
\hline\hline
\end{tabular}
\label{tab:fitparams}
\end{table}

\begin{figure}[htbp]
\centering
\includegraphics[trim={3 0 50 30},clip,width=.47\textwidth]{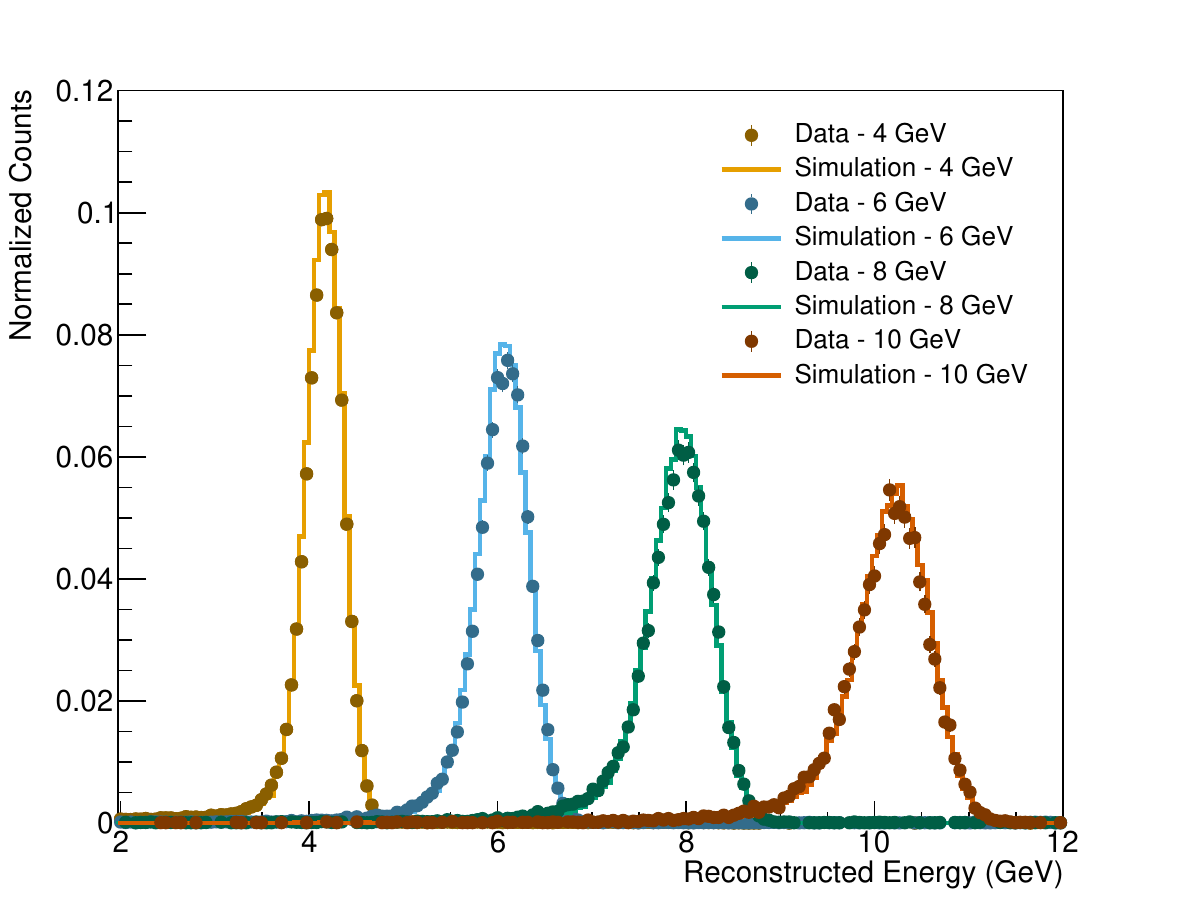}
\qquad
\includegraphics[trim={3 0 50 30},clip,width=.47\textwidth]{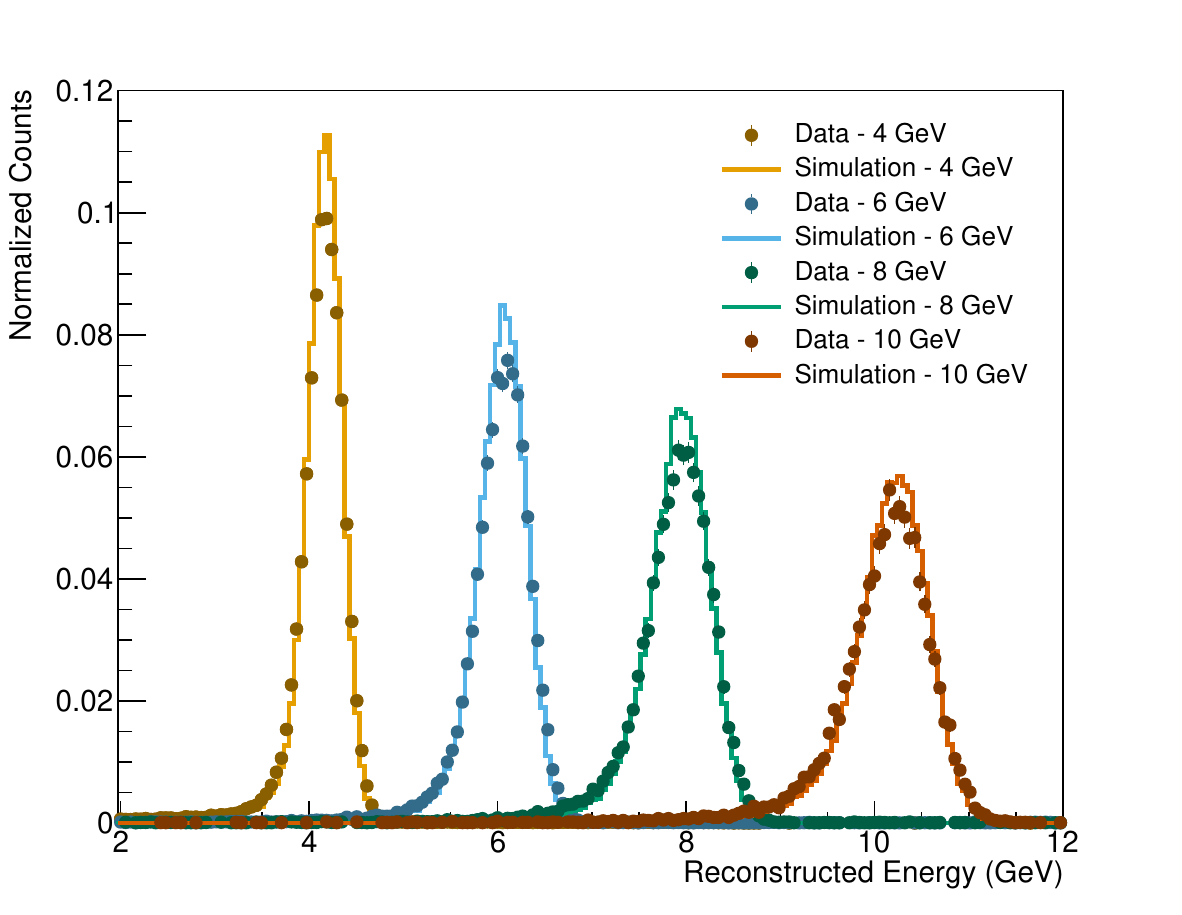}
\caption{Normalized distributions of measured and simulated electron energies at 4, 6, 8, and 10\,GeV in Baby BCAL. Left: energy distributions using the beam momentum spread ``Adjusted to the data'' values from Table~\ref{tab:beam_conditions}. Right: Same but using the ``Cited'' values. The positions of data and simulation reconstructed-energy maxima are aligned with each other to visualize the shape differences.}
\label{fig:RecoEnergiesSimu}
\end{figure}

\begin{figure}[htbp]
    \centering
    \includegraphics[trim={0 0 0 30},clip,width=1.0\textwidth]{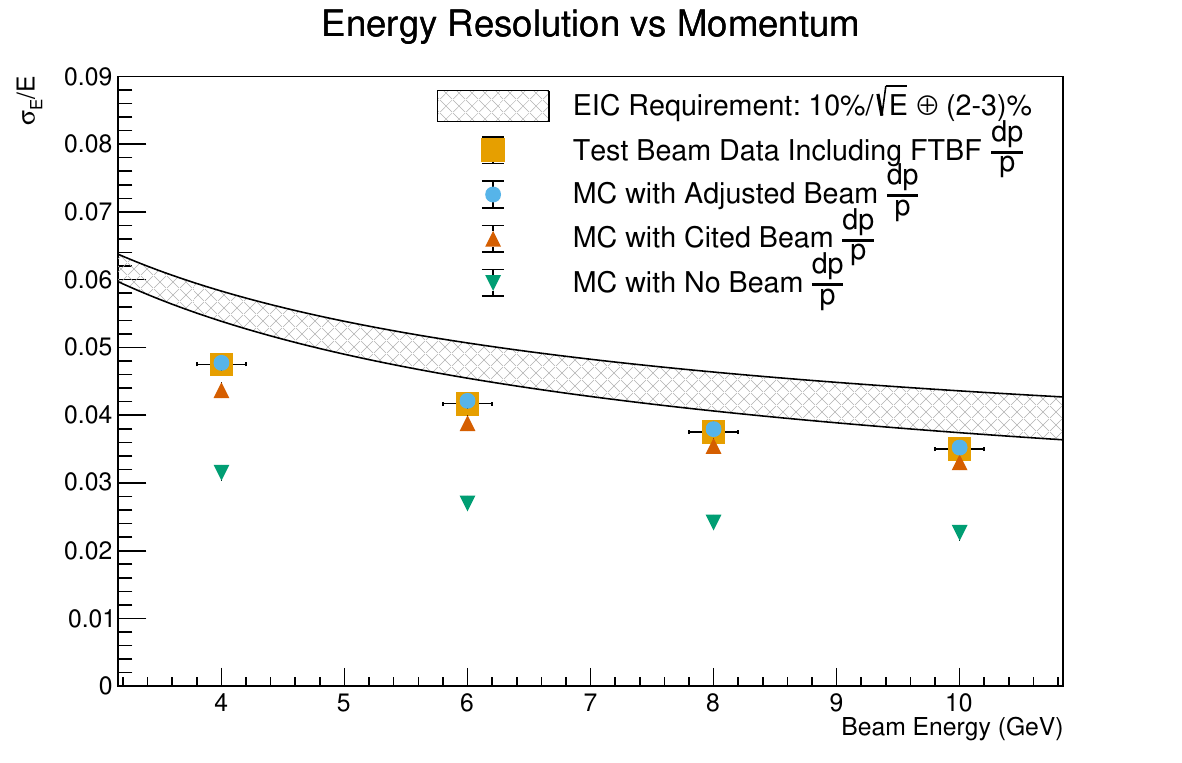}
    \caption{Measured electron energy resolution as a function of beam energy. The resolution is determined from the $\sigma$ of the Gaussian core of the Crystal Ball fits to the data and simulations. Yellow squares represent data points. Green triangles show simulations without beam momentum smearing, while red triangles include the ``Cited'' FTBF beam energy spread from Table~\ref{tab:beam_conditions}. Blue circles correspond to simulations with an adjusted beam momentum spread to match the measured data listed in the same table. The gray band represents the EIC detector system energy resolution requirement~\cite{EICRequirements}.}
    \label{fig:ERes}
\end{figure}

Figure~\ref{fig:ERes} presents the extracted energy resolution as a function of beam energy, determined from the $\sigma$ of the Gaussian core of the Crystal Ball fits to the data calibrated with the electron method and simulations. The yellow square points represent the values extracted from the data, with a 200 MeV beam energy scale uncertainty assigned. The green triangles show the values obtained from simulations without any beam momentum smearing. The red triangles correspond to the results for electrons simulated with the ``Cited'' FTBF beam momentum spread, as detailed in Table~\ref{tab:beam_conditions}, while the blue circles represent values where the beam momentum spread are adjusted to match the measured data. The gray band indicates the energy resolution requirement from the EIC Detector System Requirement database~\cite{EICRequirements}. Figure~\ref{fig:RecoEnergiesSimu} shows the reconstructed electron energy distributions from data overlaid with simulations using the two aforementioned beam momentum spread scenarios. In this plot, the positions of data and simulation reconstructed-energy maxima are aligned with each other to visualize the shape differences. For the discussion about the linearity and calibration performance see Sec.~\ref{sec:linearity}. The data and simulation shapes agree well, particularly for the adjusted beam momentum spread, with the data exhibiting slightly larger low-energy tails. This is most likely attributed to unaccounted material in front of the detector. Our simulation includes the major scintillating material present in front of the setup; however, we do not have a complete estimate of the total material thickness from all FTBF ancillary detectors. 

The extracted electron energy resolution, even when considering the \textit{full} FTBF beam energy spread, lies below the EIC detector system requirements. The simulations reproduce well the measured trends, given the uncertainties in beam conditions as the exact beam energy spread for our runs remains unknown. Due to severe time constraints during FTBF operations in the summer of 2024, dedicated runs for measuring the beam energy spread were not feasible. The actual beam momentum spread at our calorimeter prototype is expected to be larger than previously measured values cited at the FTBF wepbage (red points). This is due to the horizontal position dependence of the beam energy, introduced by the beam bending before its arrival at MTest, combined with the fact that our calorimeter is significantly longer in the horizontal direction than the lead-glass calorimeter used in past measurements. As a result, it samples a broader range of beam positions, leading to an effectively larger energy spread. This effect will be more pronounced for the 4\,GeV beam, which had a notably wide profile.

To evaluate whether the measured data can constrain the constant ($p_1$) and stochastic ($p_0$) terms of the energy resolution, we fit the data using the function $\sigma/E = p_0/\sqrt{E} \oplus p_1$ under different assumptions regarding the beam momentum spread. Since the beam momentum spread is not precisely known, we consider three scenarios: (1) the spread adjusted to the data, which is equivalent to fitting the simulation points assuming no beam momentum spread ($\Delta p/p$), (2) the cited values reported in Table~\ref{tab:beam_conditions}, where the spread varies linearly with beam energy, and (3) a constant spread of 2.3\% across all beam energies. The extracted constant term ranges from approximately 1.3\% in the adjusted and constant scenarios to about 1.9\% in the linear scenario. However, it should be noted that the fitted stochastic and constant terms are strongly anti-correlated, with a correlation coefficient of approximately -0.9. This result highlights a limitation analogous to that observed in GlueX measurements at lower energies ($<2.5$\,GeV), as discussed in the introduction, where the constant term could not be fully constrained. The data presented in this study provide a stronger constraint on the constant term, despite the beam spread uncertainty, since they are measured exclusively at higher energies ($>4$\,GeV). The fitted stochastic term varies between approximately 5.6\% for the adjusted scenario, 6.5\% for the linear scenario, and 7.7\% for the constant scenario. When fixing the stochastic term to 5.4\%, as measured for normal incidence at low energies in~\cite{Leverington:2008zz}, the fit to the adjusted scenario data yields a constant term of 1.5\%. Consequently, we conclude that the measured data constrain the constant term to be below 1.9\%.

\subsubsection{Linearity}
\label{sec:linearity}
\begin{figure}[htbp]
\centering
\includegraphics[trim={3 0 40 30},clip,width=.47\textwidth]{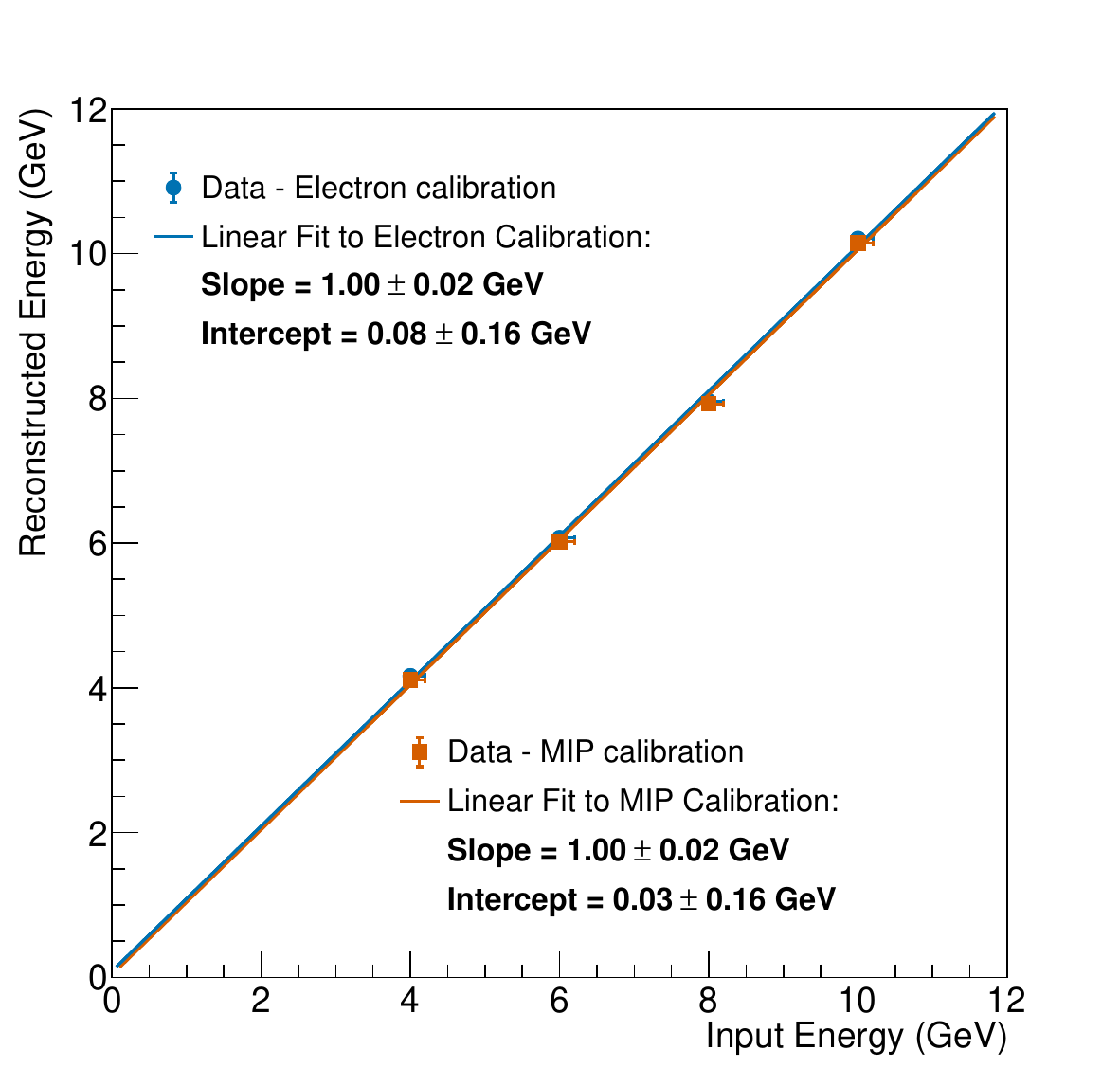}
\qquad
\includegraphics[trim={3 0 40 30},clip,width=.47\textwidth]{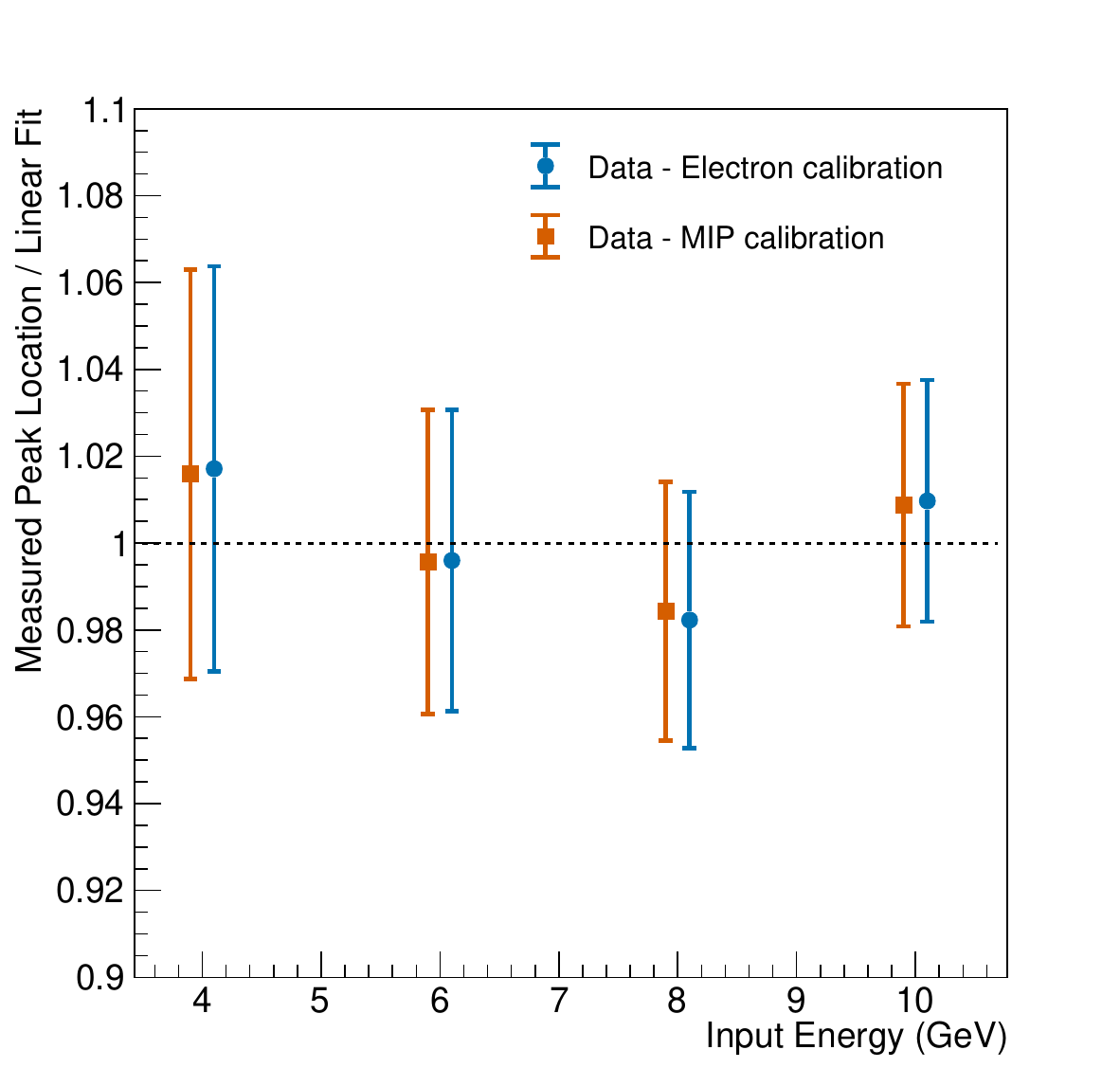}
\caption{Comparison of the two calibration methods on the linearity of the calorimeter. Left: reconstructed peak positions compared to the nominal beam energy. Right: Ratio of data points to the linear fit. The uncertainty on the data points includes the uncertainty from the fit. The MIP and electron points are offset horizontally for clarity. It can be seen that the two calibration methods are in good agreement with one another.}
\label{fig:Linearity}
\end{figure}

Using the peak locations from Fig.~\ref{fig:RecoEnergies}, the linearity of the calorimeter response is studied in Fig.~\ref{fig:Linearity}. Non-linearities can arise from various sources, including longitudinal leakage of showers or saturation of the SiPMs. Typically, such non-linearities manifest as a decrease in the amount of reconstructed energy at higher energies. Linear fits to our data without uncertainties on the input energy support the conclusion that the beam energy was approximately 200 MeV higher than requested. The MIP calibration fit results in a $y$-axis intercept of $187.8\pm1.9$ MeV, and the electron calibration fit results in an intercept of $262.9\pm2.3$ MeV. To address this, we assigned a 200 MeV one-sided uncertainty on the input energy. Fig.~\ref{fig:Linearity} (left) shows the linear fit to the reconstructed energy versus input energy dependence with the assigned one-sides uncertainty. In the case of non-linearities arising from saturating SiPM responses or leakage, the right panel of Fig.~\ref{fig:Linearity} would tend to have the 6 and 8\,GeV points above the linear fit and the 4 and 10\,GeV points below it. We therefore conclude that with the four energy points measured, we do not observe non-linearity within our uncertainties.

\subsection{Pion Energy Response}
\label{sec:PionEnergyResponse}
\begin{figure}[htbp]
\centering
\includegraphics[trim={3 5 55 30},clip,width=.47\textwidth]{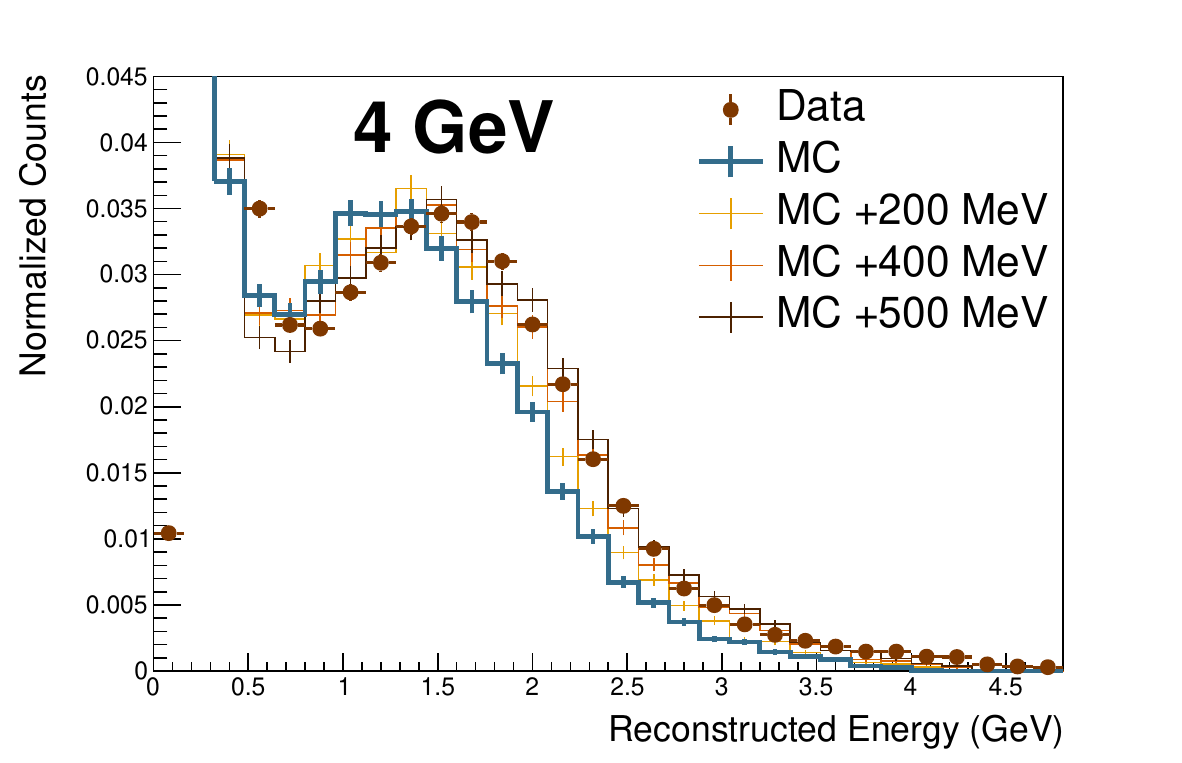}
\qquad
\includegraphics[trim={3 5 55 30},clip,width=.47\textwidth]{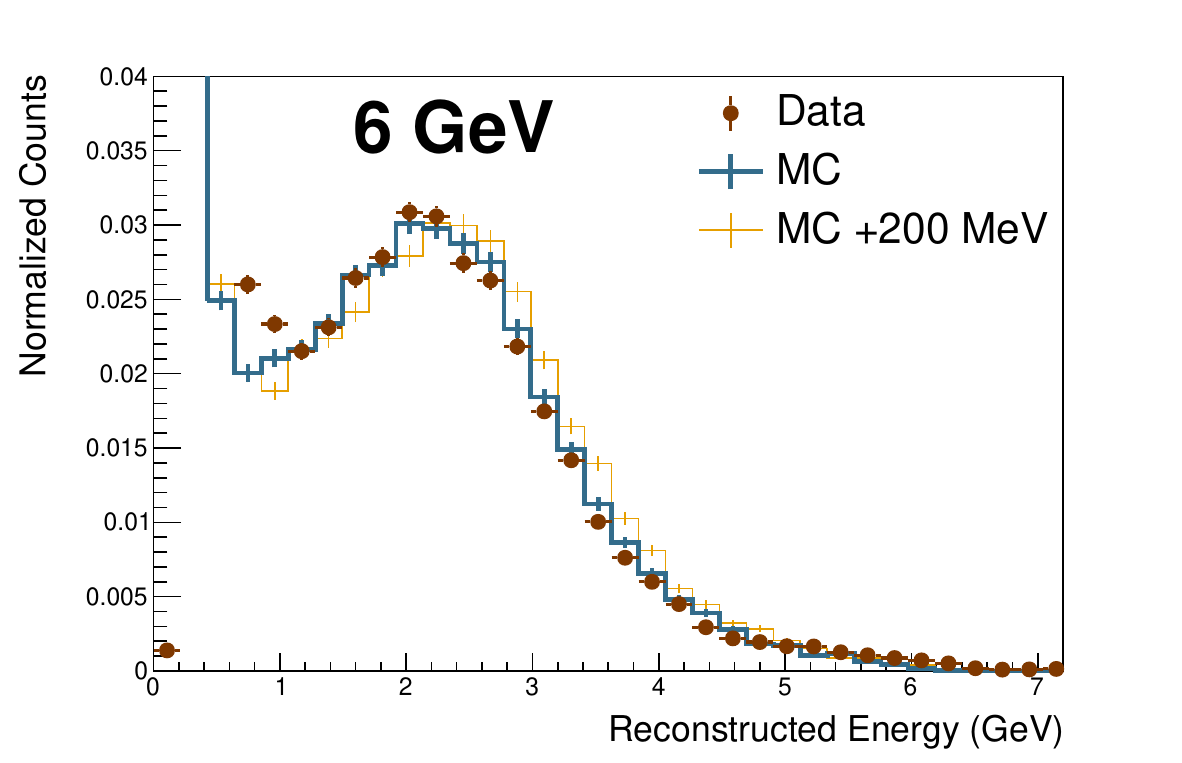}
\qquad
\includegraphics[trim={3 5 55 15},clip,width=.49\textwidth]{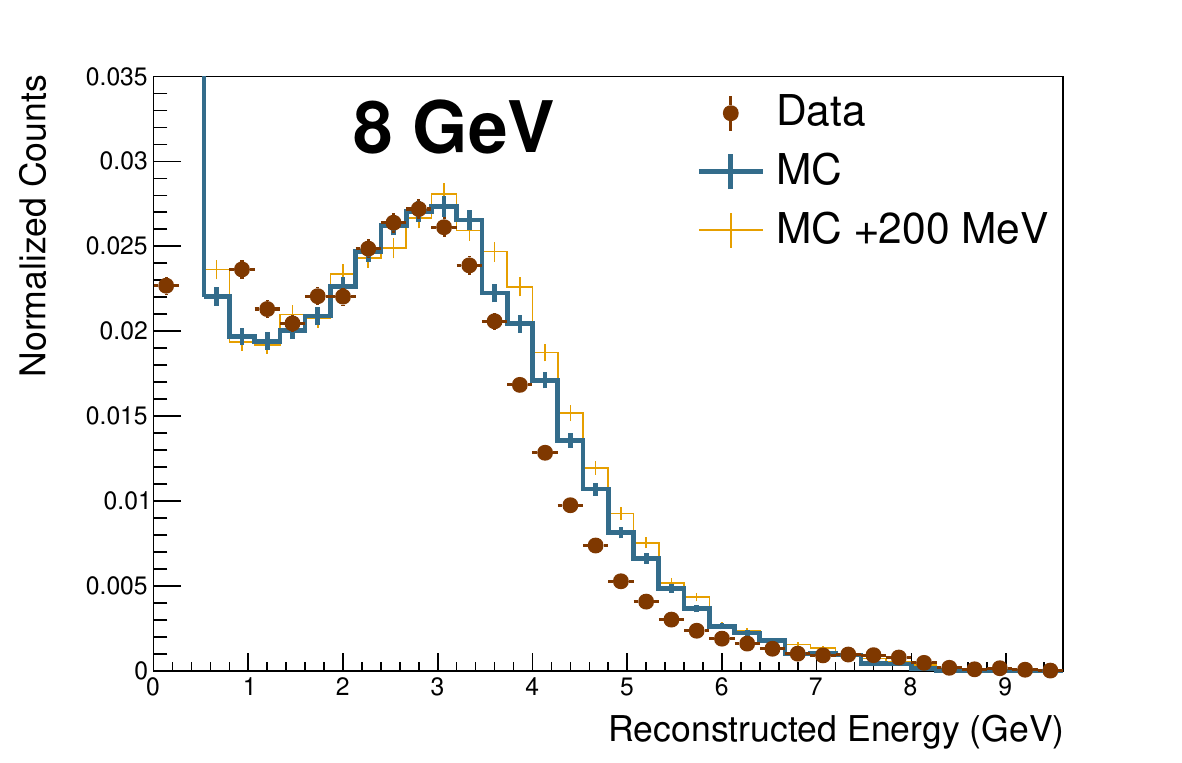}
\caption{Pion energy response for 4\,GeV, 6\,GeV, and 8\,GeV data. Circular markers represent data, calibrated using the electron method. The yellow and blue histograms correspond to simulations with nominal and $+200$\,MeV shifted energies, respectively, using the ``Adjusted to the data'' beam momentum spread from Table~\ref{tab:beam_conditions}. Simulations are normalized by the height the peak of the energy loss spectrum for showering pions.}
\label{fig:pionSpectrum}
\end{figure}

We characterize the pion energy response and present the results in Fig.~\ref{fig:pionSpectrum} for 4\,GeV, 6\,GeV, and 8\,GeV data. The 10\,GeV data were taken during a period when the calorimeter channel thresholds, due to technical issues, did not allow for efficient detection of minimum ionizing particles or low-energy pion deposits, preventing a straightforward comparison between data and simulations. The circular markers represent data for events identified as pions. The yellow histogram corresponds to simulations for pions with nominal energy, while the blue histogram represents simulations with a nominal energy $+200$\,MeV. For the simulations we assumed the ``Adjusted to the data'' beam momentum spread, as shown in Table~\ref{tab:beam_conditions}. We normalize the simulated distributions to the data by scaling them so that the peak height of the energy loss spectrum for showering (non-MIP) pions matches between data and simulation. To achieve this, we tested several fitting functions for both data and simulation spectra in the region of the showering pion peak, including Crystal Ball and Gaussian functions combined with first- to third-order polynomials, over energy ranges appropriate for each beam energy. The final normalization was based on the peak height extracted from a Gaussian plus first-order polynomial fit, which provided a stable and consistent description across datasets. 

Overall, the agreement between data and simulation is reasonable. However, in the case of the 8\,GeV data, the simulation is slightly shifted toward higher reconstructed energies, with the fitted peak position offset by about 100\,MeV relative to the nominal energy simulation, and the right-side shoulder slightly overestimating the data. At 4\,GeV, this discrepancy is more pronounced, with the simulation shifted toward lower reconstructed energies and the fitted peak position offset by about 270\,MeV.  

To investigate potential sources of these discrepancies, we examined the impact of the physics list used in the simulations. The results in Fig.~\ref{fig:pionSpectrum} are generated using the FTFP\_BERT physics list. Additionally, we tested the QGSP\_BERT, QGSP\_BERT\_HP, and FTFP\_BERT\_HP physics lists. As shown in Fig.~\ref{fig:pionSpectrumList}, no significant differences were observed between the pion energy responses across these physics lists, beyond the statistical uncertainties of the generated samples.

Another possible explanation for the observed shifts might be the modeling of hadronic signal quenching, specifically the Birks' constant implemented in our simulation. The effect of Birks' constant on the pion reconstructed energy is illustrated in Fig.~\ref{fig:pionSpectrumkB}. To account for the observed data-simulation shift at 4\,GeV, Birks' constant would need to be reduced to approximately $25\%-50\%$ of its nominal value of 0.132\,mm/MeV. However, such a large modification would be inconsistent with the literature value, which was specifically measured for the type of fibers used in Baby BCAL~\cite{POSCHL2021164865}. Furthermore, as shown in Fig.~\ref{fig:pionSpectrumkB}, such a change would lead to an overestimation of deposited energies in the 6\,GeV and 8\,GeV pion data.

To assess the impact of beam conditions, we tested variations in the beam profile by shifting the beam width at the calorimeter face within one sigma of the extracted values (see Sec.~\ref{sec:FTBF}). However, no significant difference was observed in the reconstructed pion energy, as shown in Fig.~\ref{fig:pionSpectrumProfile}.

Finally, we examined whether an offset in beam energy could account for the data-simulation shift at 4\,GeV. As seen in Fig.~\ref{fig:pionSpectrum}, a shift of approximately $+400-500$\,MeV would be required to reconcile the data and simulation. We noticed, that the 4\,GeV beam profile was particularly irregular, which may have impacted the beam momentum spread. Unfortunately, we do not have a direct way to confirm the full $+400-500$\,MeV shift. The assumption of a $+200$\,MeV to $+500$\,MeV beam-momentum-mean uncertainty would be supported by our calibration and linearity check (see Sec.~\ref{sec:datacalib}), however, a $+400-500$\,MeV shift at 4\,GeV would not preserve a similar level of linearity as presented in Fig.~\ref{fig:Linearity}.

To summarize: The overall level of agreement between data and simulation is reasonable, with some discrepancies observed, particularly at 4\,GeV. We explored several potential explanations for these differences. Variations in the physics list and beam profile spread had no significant impact. Adjusting Birks' constant to match the 4\,GeV data would require an unreasonably small value, inconsistent with literature, and lead to inconsistencies at higher energies. A possible contributor to the observed agreement between data and simulation is the mean beam momentum and its spread. The relative shift between data and simulation is consistent with the results from the calibrated electron data in Fig.~\ref{fig:RecoEnergies} and Table~\ref{tab:fitparams}: the 4\,GeV data overshoots, the 6\,GeV data agrees with, and the 8\,GeV data undershoots the nominal beam energy value. However, a direct confirmation is not possible with the available dataset. Given the constraints of the FTBF dataset, this level of agreement represents the best achievable precision. Further improvements would require dedicated long runs with reference beam energy profile measurements after each tuning step. Such measurements were not feasible in this test campaign due to severe beam time limitations in 2024.

\subsection{Separation of Electrons from Pions}
\label{sec:episeparation}
Showers initiated by electrons and pions can be separated based on the shower shapes and/or the total amount of energy deposited in the calorimeter. We first characterize the capability of the Baby BCAL to do $e/\pi$ separation in the same way as a standard tower-based calorimeter using the total energy measurement, which we denote as the $E/p$ method, before moving on to a machine learning method that includes shower shape information. We define the pion rejection factor $R_{\pi}$ as the total number of pions incident on the detector divided by the number surviving after electron selection cuts have been applied. Unless otherwise specified, $R_{\pi}$ is quoted at the point where 95\% of electrons survive the electron selection cuts. Pion showers typically deposit less than half of the incident beam energy, with a tail to higher energies corresponding to pions which shower early and primarily electromagnetically. 

\begin{figure}[h!]
\centering
\includegraphics[trim={3 0 40 30},clip,width=.47\textwidth]{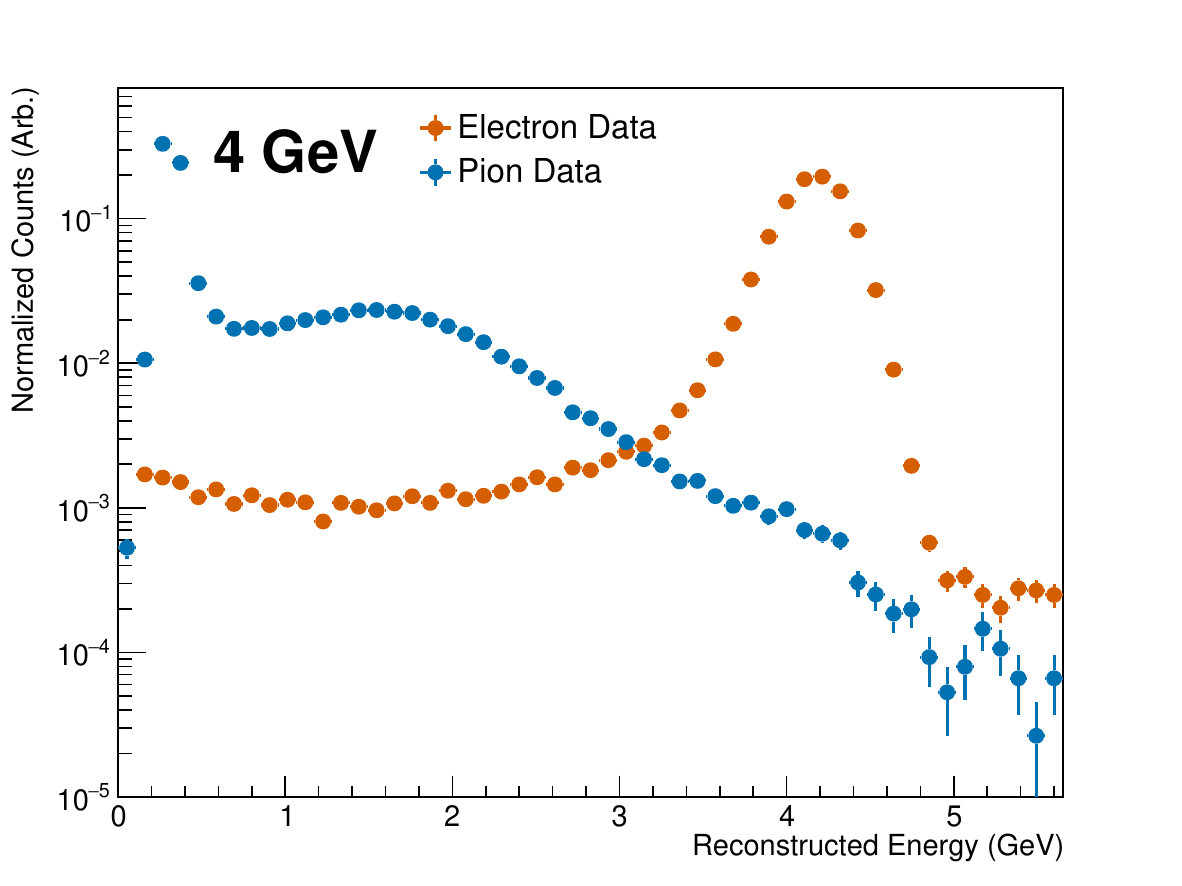}
\includegraphics[trim={3 0 40 30},clip,width=.47\textwidth]{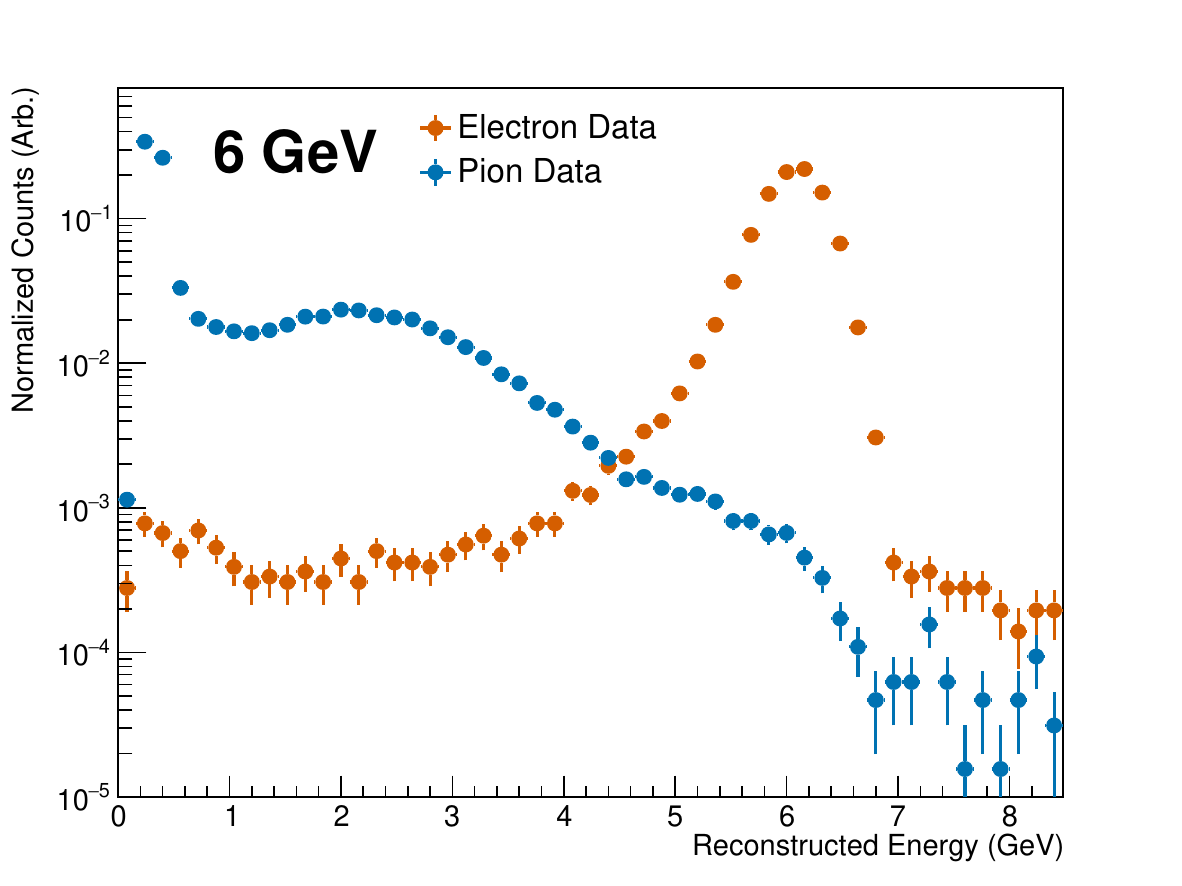}
\includegraphics[trim={3 0 40 30},clip,width=.47\textwidth]{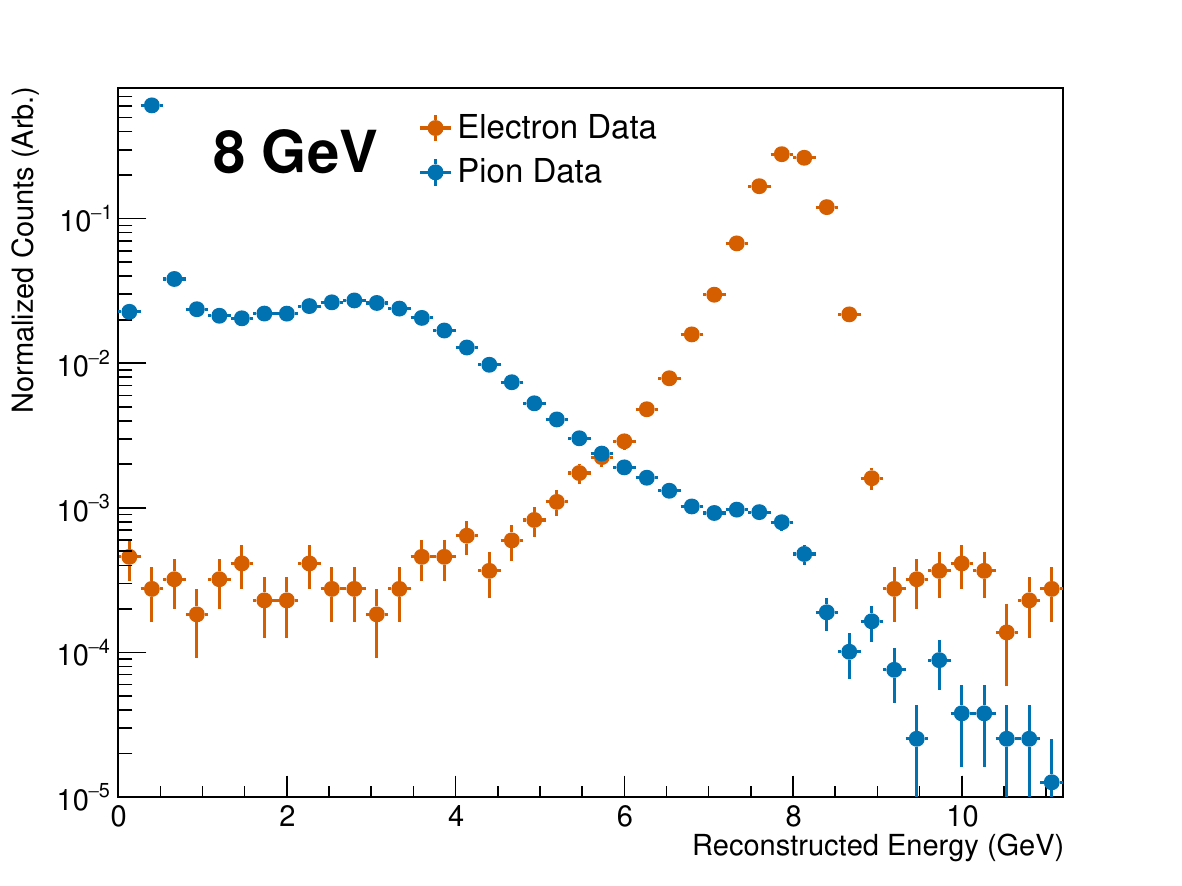}
\includegraphics[trim={3 0 40 30},clip,width=.47\textwidth]{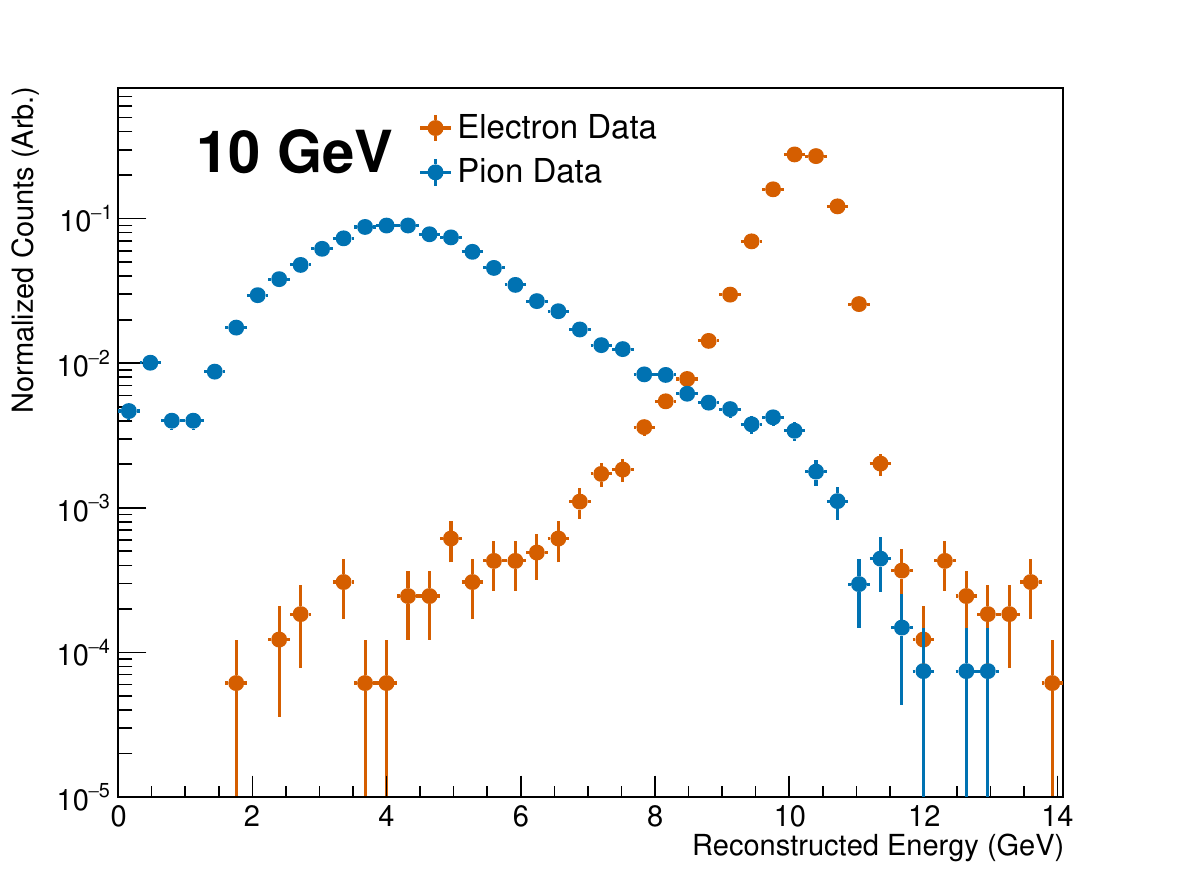}
\caption{Comparison of electron and pion data using the electron calibration method at the four beam energies in log-scale. Note the changing x-axes.}
\label{fig:PionData}
\end{figure}

An important factor in quantifying the total $e/\pi$ separation performance is the total number of pions which pass through the calorimeter as minimum ionizing particles without showering. Although these particles will realistically never be confused for electrons, understanding their number is necessary to provide a numerical value for the $\pi$ rejection performance. Comparing the data and simulation of our setup, the number of pions depositing less than 1\,GeV of energy in the calorimeter was observed to match within a couple percent for all energies except for the 10\,GeV data. The 10\,GeV data were taken during a period where the thresholds of the calorimeter channels did not permit efficient detection of MIPs. For the 10\,GeV data, we utilize the simulation to understand the fraction of pions that are expected to interact only as MIPs. For the most relevant region for $\pi$ rejection, namely pions leaving high energy deposits, we use the pion data collected. This pion data is shown in Fig.~\ref{fig:PionData}. 

Several features relevant to the extraction of pion rejection performance should be noted in Fig.~\ref{fig:PionData}. The Cherenkov-selected pion datasets exhibit a peak at the nominal beam energy. This peak could be the result of electrons which did not fire the Cherenkov detectors and thus were misidentified as pions. Given the previously mentioned Cherenkov detector efficiencies, the expectation is that only one electron in one million would accidentally not fire the Cherenkov detectors. One possibility is that these events result from pileup of two showering pion events in the same data acquisition window, however the sharpness of the peak, particularly in the 8\,GeV data, seems to favor the electron misidentification hypothesis. A small number of events with energies greater than the beam energy are observed in both the pion and the electron samples arising from pileup.

Furthermore, there is a tail of Cherenkov-selected electron events which extends down to very low energies. This effectively lowers $R_{\pi}$ because the threshold value of $E/p$ for maintaining 95\% electron efficiency must be lowered. The fact that the tail is strongest at lower momenta suggests interactions with material upstream of the Baby BCAL could be responsible. Another factor could be the spatial spread of the beam, which is larger at lower momenta. Due to the limited vertical extent of the Baby BCAL, a larger vertical spread of the beam would produce a tail to lower energies from electrons which impact only on the edge of the calorimeter. However, the ``U'' cut (described in Sec.~\ref{sec:channel-level-analysis-cuts}) applied to all events in this dataset requires particles to have their highest energy deposits in the central channels of the calorimeter, mitigating the otherwise deleterious effects of transverse electron shower leakage. The 4 and 6\,GeV electron data clearly shows a peak at very low energy, possibly from radiated photons emitted by electrons that missed the calorimeter face or showers in material upstream of the Baby BCAL. This tail is an artifact of the present test setup and is not indicative of the intrinsic detector response. In a realistic experiment, the strength of this tail will depend strongly on the amount of material traversed by particles before impacting the calorimeter. The composition of the dataset used in the evaluation of the $e/\pi$ separation in this analysis is presented in Table~\ref{tab:dataset_composition}. 

\begin{table}[htb]
    \centering
    \caption{Composition of the dataset used in the evaluation of the $e/\pi$ separation. The number of Cherenkov-selected electrons and pions at each beam energy is listed.}
    \begin{tabular}{lcc}
        \hline\hline
        Beam energy (GeV) & Electrons & Pions \\
        \hline
        4  & 136999 & 68277 \\
        6  & 42877  & 58166 \\
        8  & 27988  & 72022 \\
        10 & 21571  & 13374 \\
        \hline\hline
    \end{tabular}
    \label{tab:dataset_composition}
\end{table}

To partially isolate the intrinsic response of the detector, we compare two fits to the measured electron distribution with our measured pion distribution. To eliminate the very low energy electron events we use both a Gaussian and a Crystal Ball fit that includes only reconstructed energies greater than half the nominal beam energy. The Gaussian fit naturally provides better separation due to the absence of a tail, representing the pion rejection performance under the idealized conditions for electrons with no upstream material and full shower containment\footnote{However, the Gaussian width and high-energy pion showers are still affected by the FTBF beam momentum spread (see Sec.~\ref{sec:energyres}), resulting in less-than-idealized $e/\pi$ separation.}. Figure 11.48 of the EIC Yellow Report~\cite{YellowReport}, which summarizes world results for $e/\pi$ separation, also assumes this approximation. However, any comparison with our results must take into account that both the Gaussian and Crystal Ball fits in our analysis are applied to the measured electron data and therefore include the deleterious effect of the FTBF beam momentum spread. 

\begin{figure}[htbp]
\centering
\includegraphics[width=0.9\textwidth]{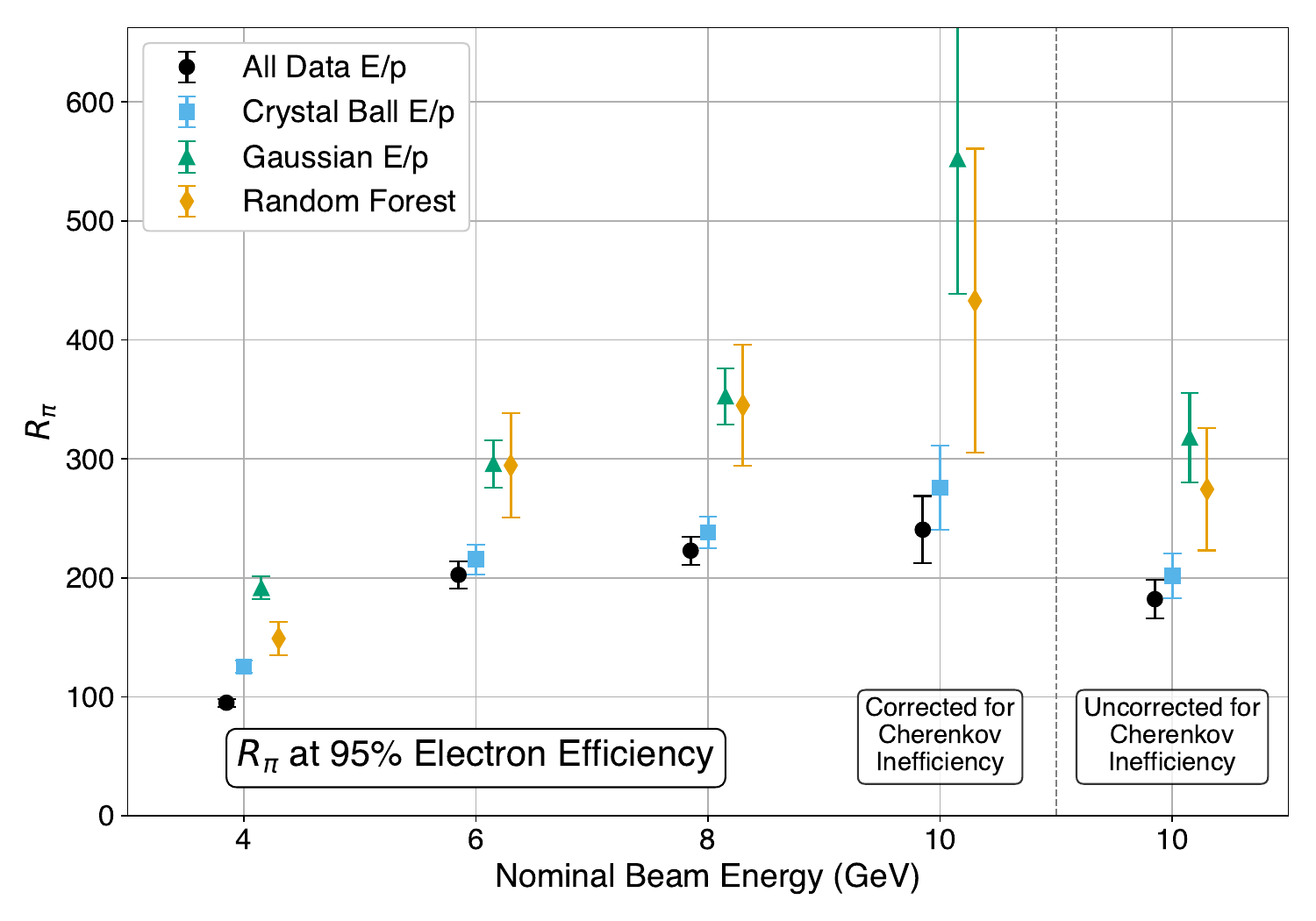}
\caption{Measured pion rejection factor $R_{\pi}$ as a function of beam energy at 95\% electron efficiency, determined using different electron-pion separation methods: the ``All Data $E/p$'' (black circles), ``Gaussian $E/p$'' (green triangles), ``Crystal Ball $E/p$'' (blue squares), and ``Random Forest'' (yellow diamonds) under the test beam conditions. The points are offset horizontally for clarity. These methods, in various degree, incorporate artifacts of the test beam setup, including beam momentum smearing, effects of upstream material, possible pileup, and beam profile variations. Readers should refer to the text for a detailed discussion of the assumptions and limitations of each method.}

\label{fig:epiSeparation}
\end{figure}

The results of this study are presented in Fig.~\ref{fig:epiSeparation} showing the pion rejection factor $R_{\pi}$ as a function of beam momenta. The uncertainties correspond to the statistical uncertainty arising from the number of pions surviving the cut. In the case of the Gaussian method (green triangles), where the cut on $E/p$ to achieve 95\% electron efficiency is quite strict, this uncertainty becomes large due to the small number of pions surviving. Black circles mark $R_{\pi}$ values calculated with $E/p$ method for all registered data presented in Fig.~\ref{fig:PionData}. It can be seen from the figure that the Crystal Ball method (blue squares) only significantly outperforms the all data for the 4\,GeV point due to the larger tail to lower energies observed at that energy. 

We also study the ability of a random forest classifier to separate electrons from pions based on only the uncalibrated signal amplitudes in ADC units. Events passing the global event cuts, namely the ``U'' cut and the 500 MeV energy difference cut described in Sec.~\ref{sec:channel-level-analysis-cuts}, are sorted into pion and electron training samples based on the information from the Cherenkov detectors. Crucially, this includes the tails of the electron distributions seen in Fig.~\ref{fig:PionData}. The random forest method trains a large number of neural network classifiers on subsets of the data, then classifies an event as an electron or a pion based on the majority vote of the networks. For the results presented here, 100 classifiers are trained and used to predict the particle type. 20\% of the dataset is retained for testing the performance of the network. To quantify the uncertainty on the $R_{\pi}$ result, the procedure of training and classifying with the 100 networks is repeated 200 times with random subsets of the events being sent to each network for training. The standard deviation of the pion rejection factors determined in the 200 repetitions is taken as the systematic uncertainty on $R_{\pi}$ associated with this method. The statistical uncertainty arising from the number of pions surviving the electron selection cuts, typically on the order of a couple hundred, is simply added in quadrature with the random forest model uncertainty despite the fact that these two uncertainties are technically correlated. At all energies the random forest systematic uncertainty dominates. It can be seen from Fig.~\ref{fig:epiSeparation} (yellow diamonds) that the random forest technique performs similarly to the Gaussian $E/p$ method. This may seem surprising since the inclusion of shower shape information typically provides significant benefits compared to $E/p$ alone. However, the random forest technique has to contend with the low-energy tail of the electron data while the Gaussian $E/p$ does not. The fact that the random forest performs similarly to the Gaussian $E/p$ suggests that the low-energy tail can be appropriately classified by the random forest.

It was noticed during the analysis that the value of $R_{\pi}$ appeared to drop for all methods at 10\,GeV compared to 8\,GeV, while the naive expectation is that $R_{\pi}$ should continue to grow or remain constant with energy. This feature was identified as related to mis-identification of electrons by the Cherenkov detectors, which were operated at different pressures and with different thresholds on their analog signals during the 10\,GeV data collection period. The Cherenkov detector inefficiency in the 10\,GeV data is such that one in every 1150 electrons will be misidentified as a pion, to be compared to the one in every one million for the other datasets. The 10\,GeV dataset after all analysis cuts consists of 21537 Cherenkov-identified electrons and 13493 Cherenkov-identified pions. When weighted with the relative numbers of pions and electrons, the reduced Cherenkov detector efficiency means that the best achievable $R_{\pi}$ given this dataset is around 600. For the $E/p$ cut using just the electron and pion data (black circles in Fig.~\ref{fig:epiSeparation}), 74 Cherenkov-selected pion events survive the $E/p$ cut, which should be compared with the expected 18 electrons mis-identified as pions and passing the electron selection cuts. We can estimate the true $R_{\pi}$ by subtracting the expected number of electrons mis-identified as pions from the number of pions surviving. Since the 10\,GeV data is the least statistically precise, the constraint on the value of $R_{\pi}$ is weaker than the other energies, but the data qualitatively agree with the expected trend within uncertainties when corrected for Cherenkov detector inefficiency.

The results of this study provide insights into the electron-pion separation capabilities of the Baby BCAL under test beam conditions. The measured pion rejection factors include the performance degradation due to beam energy spread, upstream material interactions, pileup, and beam profile variations that cannot easily be unfolded to obtain the ``true'' detector performance under ideal conditions. The Gaussian method yields rejection factors that represent the performance achievable with the $E/p$ technique without upstream material and leakage, though still affected by the momentum spread’s broadening effect. With the expected energy resolution corrected for beam effects, the performance will improve further.

Additionally, the results highlight the potential of using longitudinal shower profile information for electron-pion discrimination. The random forest method, trained on uncalibrated ADC values, achieved similar rejection factors to the Gaussian method but for the \textit{entire} measured dataset, including events in the low-energy electron tail. Incorporating shower shape information into electron-pion separation, even with the limited segmentation of the Baby BCAL, improved pion rejection by a factor of approximately 1.6–1.8, depending on beam energy. This suggests that leveraging longitudinal shower profile features helps mitigate the impact of low-energy electron tails in the test beam environment and improves separation performance. In the full-scale detector during real operations, these effects are expected to be more pronounced, further enhancing electron-pion separation in the Pb/ScFi section of BIC. This improvement will stem from the finer longitudinal segmentation, increased calorimeter depth, and the absence of FTBF beam-related artifacts.

\section{Summary and Outlook}

The Baby BCAL was successfully tested to assess its response to electrons and pions. As a small-scale Pb/ScFi prototype of the GlueX BCAL, the Baby BCAL serves as a benchmark for the Pb/ScFi part of the future BIC in the ePIC detector at the EIC. Compared to the planned BIC, the Baby BCAL is shallower, with 15.5 radiation lengths at normal incidence versus 17.1, and has less segmentation, providing a simplified testbed for evaluating detector performance and reconstruction techniques. The test beam was performed with electrons and pions at energies ranging from 4 to 10\,GeV, extending previous GlueX studies to higher energies relevant for EIC applications.  

The detector exhibited good linearity within uncertainties. Despite the fact that the electron energy resolution was substantially affected by the FTBF beam momentum spread (of the order of 2–3\%), the measured resolution still meets the EIC requirements for being better than $10\%/\sqrt{E} \oplus (2\text{--}3)\%$. By analyzing different beam momentum spread scenarios, the measured data further constrain the constant term in the energy resolution to be below 1.9\% for particles at normal incidence, improving previous constraints obtained at lower energies. Simulations describe the data reasonably well for both electrons and pions given the constraints of the test beam environment, increasing confidence in their use for modeling the full-scale BIC. A limiting factor of this study, however, was the lack of precise knowledge of the beam energy mean and its spread and the limited sensitivity to the stochastic term of the energy resolution within this measurement.

We studied electron-pion separation performance using multiple methods based on $E/p$ and a random forest classifier, incorporating varying degrees of test beam effects. The measured pion rejection factors at 95\% electron efficiency, reaching up to approximately 550 depending on beam energy, reflect the combined effects of the detector response and the full test beam environment, including upstream material, beam energy and spatial spread, and possible pileup. This study demonstrates that incorporating longitudinal shower profile information, even with the limited readout granularity of the Baby BCAL, plays an important role in electron-pion separation.

These results establish a basis for further improving detector performance, refining simulation and reconstruction models, and guiding future beam tests. A separate, ongoing analysis of positron test beam data taken in Hall D at Jefferson Lab in 2023, where the beam energy spread is significantly lower, will offer a complementary perspective on the detector’s electromagnetic response at energies up to 6\,GeV. Additionally, potential future measurements, particularly with pions, could further refine the experimental setup and expand the scope of detector characterization. If additional test beams are conducted, efforts will focus on optimizing the delivered beam profile, precisely measuring the beam energy and its spread for each run, and utilizing tracking detectors to fully characterize the beam profile. These enhancements, which were not feasible during the 2024 campaign due to limited running time, present an opportunity to further characterize and validate the intrinsic performance of the tested prototype.

Building on these results, further efforts will explore refined reconstruction techniques, particularly leveraging shower profile information for enhanced electron-pion separation in the ePIC environment. The insights gained from this test beam will play a key role in guiding final design choices for the BIC, including optimizing detector readout, improving energy calibration strategies, and refining the simulations to support high-precision electromagnetic calorimetry at the EIC.  

\acknowledgments
This document was prepared by the members of the Barrel Imaging Calorimeter (BIC) Detector Subsystem Collaboration (DSC) of the ePIC Collaboration using the resources of the Fermi National Accelerator Laboratory (Fermilab), a U.S. Department of Energy, Office of Science, Office of High Energy Physics HEP User Facility. Fermilab is managed by FermiForward Discovery Group, LLC, acting under Contract No. 89243024CSC000002. The material is based upon work partially supported by the U.S. Department of Energy, Office of Science, Office of Nuclear Physics and Laboratory Directed Research and Development (LDRD) funding from Argonne National Laboratory, provided by the Director, Office of Science, of the U.S. Department of Energy under Contract No. DE-AC02-06CH11357; the Electron-Ion Collider Project R\&D Funds for the eRD115 Project; and Natural Sciences and Engineering Research Council of Canada Grant No. SAPPJ-2023-00041.

We thank the ePIC Software and Computing Working Group and all contributors for maintaining the ePIC software, which was essential for our simulations and analysis. We also acknowledge the invaluable support of the FTBF staff, whose expertise and dedication greatly facilitated the execution of our test beam experiments. Additionally, we express our gratitude to the Hall D staff and the GlueX Collaboration for lending the Baby BCAL for these tests, enabling this study to be carried out successfully.

\appendix
\label{app:EventDisplay}
\section{Example Event Displays}
\begin{figure}[htbp]
\centering
\includegraphics[trim={125 50 95 85},clip,width=0.99\textwidth]{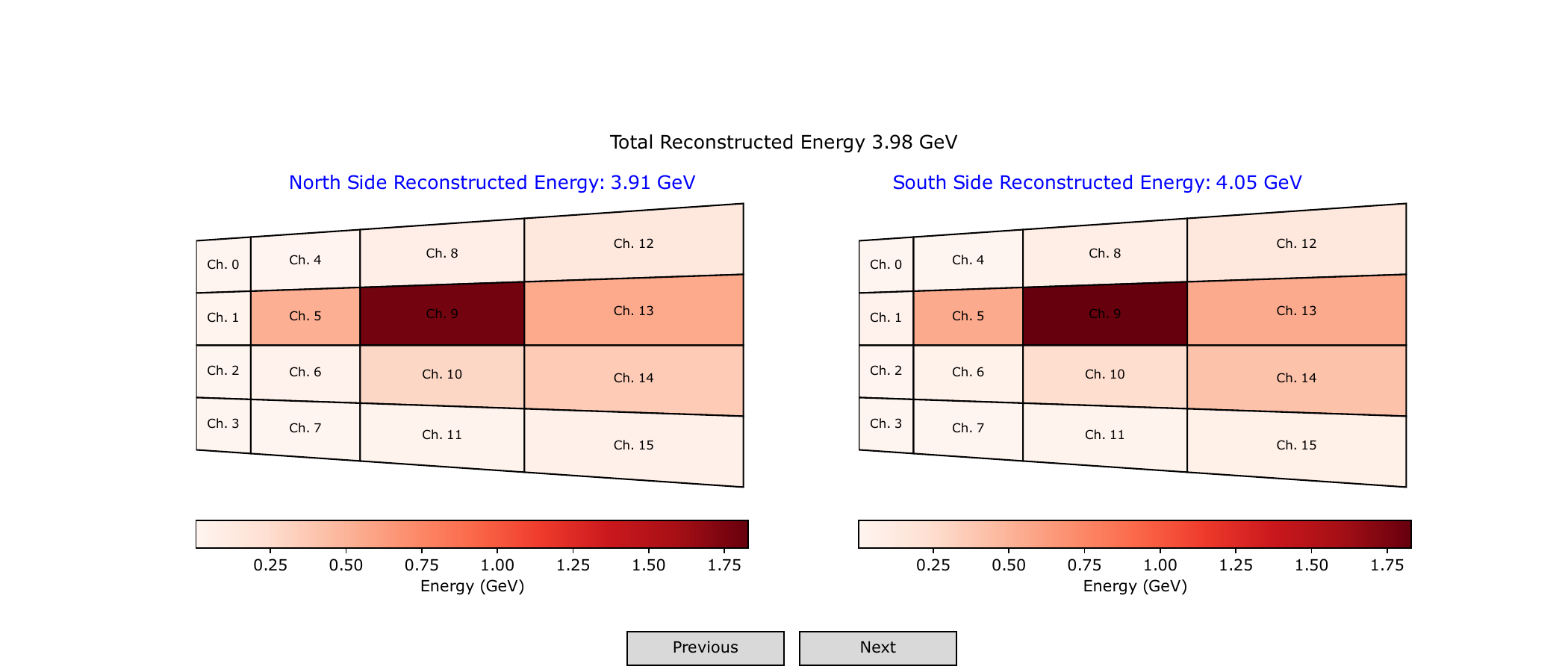}
\caption{Example event display for 4\,GeV electron from the FTBF beam identified with the Cherenkov detectors. The calibrated energy response in each channel on both south and north side of the Baby BCAL has been presented. On the bottom the energy scale is presented.}
\label{fig:EventDisplay4GeVElectron}
\end{figure}

\begin{figure}[htbp]
\centering
\includegraphics[trim={125 50 95 85},clip,width=0.99\textwidth]{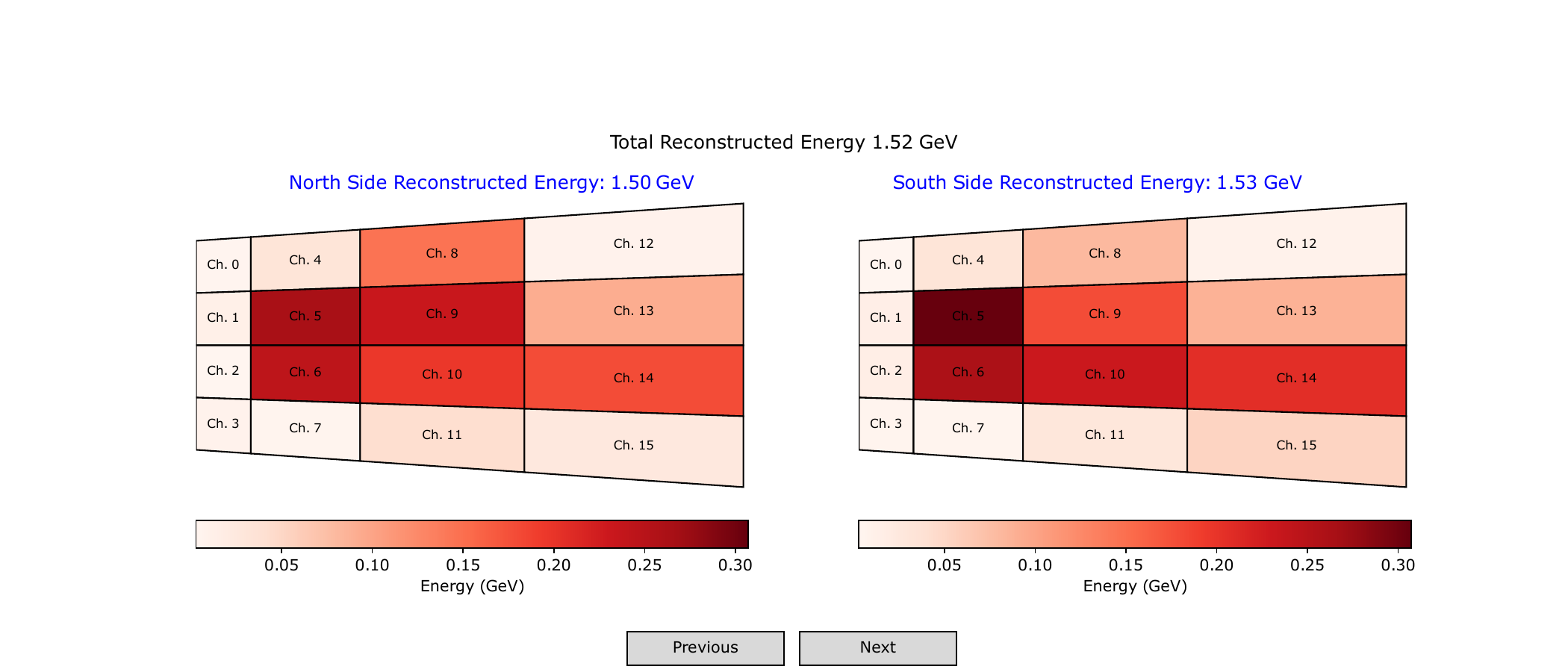}
\qquad
\includegraphics[trim={125 50 95 85},clip,width=0.99\textwidth]{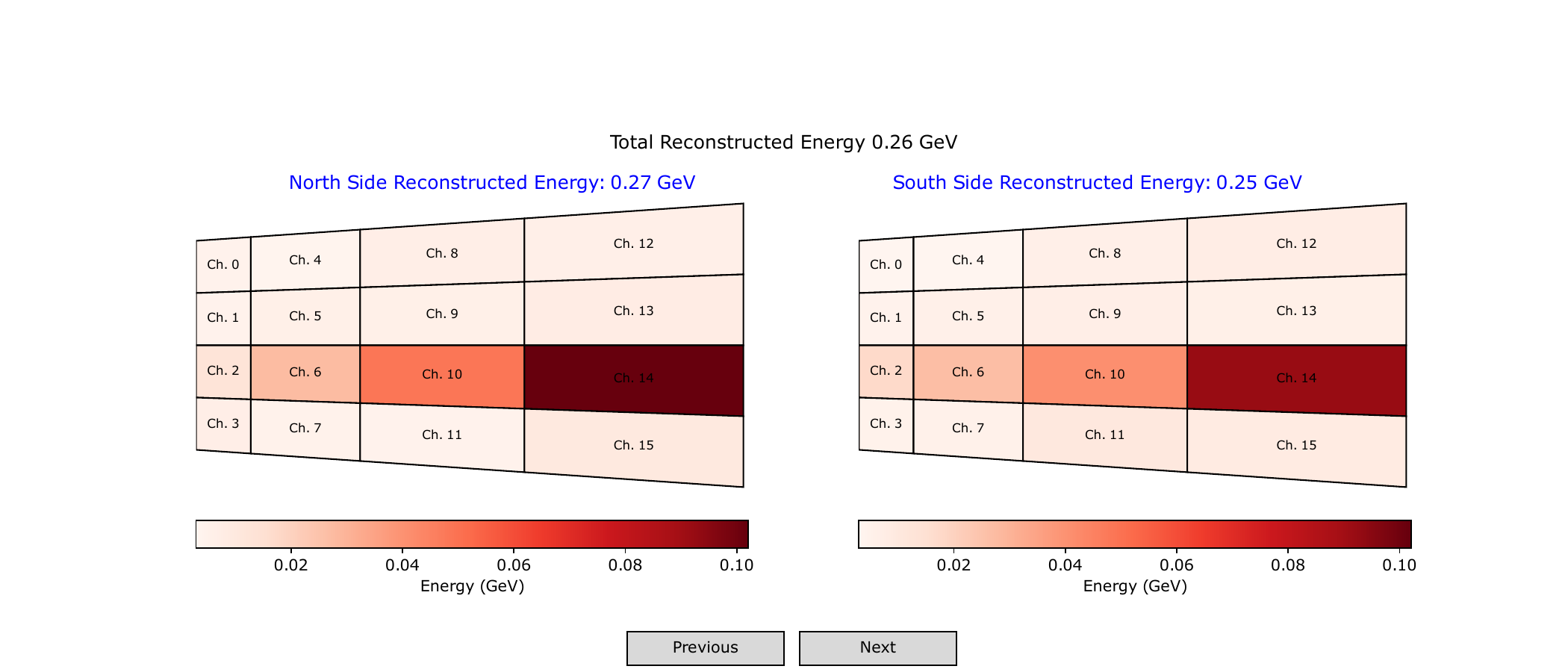}
\caption{Example event display for 4\,GeV pion from the FTBF beam identified with the Cherenkov detectors. The calibrated energy response in each channel on both south and north side of the Baby BCAL has been presented. The top plots show a pion that initiated a shower in the Baby BCAL. The bottom plots show a pion that leave MIP-like signal depositing energy in one distinct column of the Baby BCAL.}
\label{fig:EventDisplay4GeVPionShower}
\end{figure}

This appendix presents example event displays showing the calibrated energy response in the channels of the Baby BCAL. These displays illustrate the differences in energy response topologies for various particles: an electron (Fig.~\ref{fig:EventDisplay4GeVElectron}), a pion initiating a shower in the baby BCAL and a pion leaving a MIP-like signal (Fig.~\ref{fig:EventDisplay4GeVPionShower}). The particles have been identified with the Cherenkov detectors as described in Sec.~\ref{sec:PID}.

\clearpage
\section{Pion Response in Simulations}
\begin{figure}[htbp]
\centering
\includegraphics[trim={3 5 55 30},clip,width=.47\textwidth]{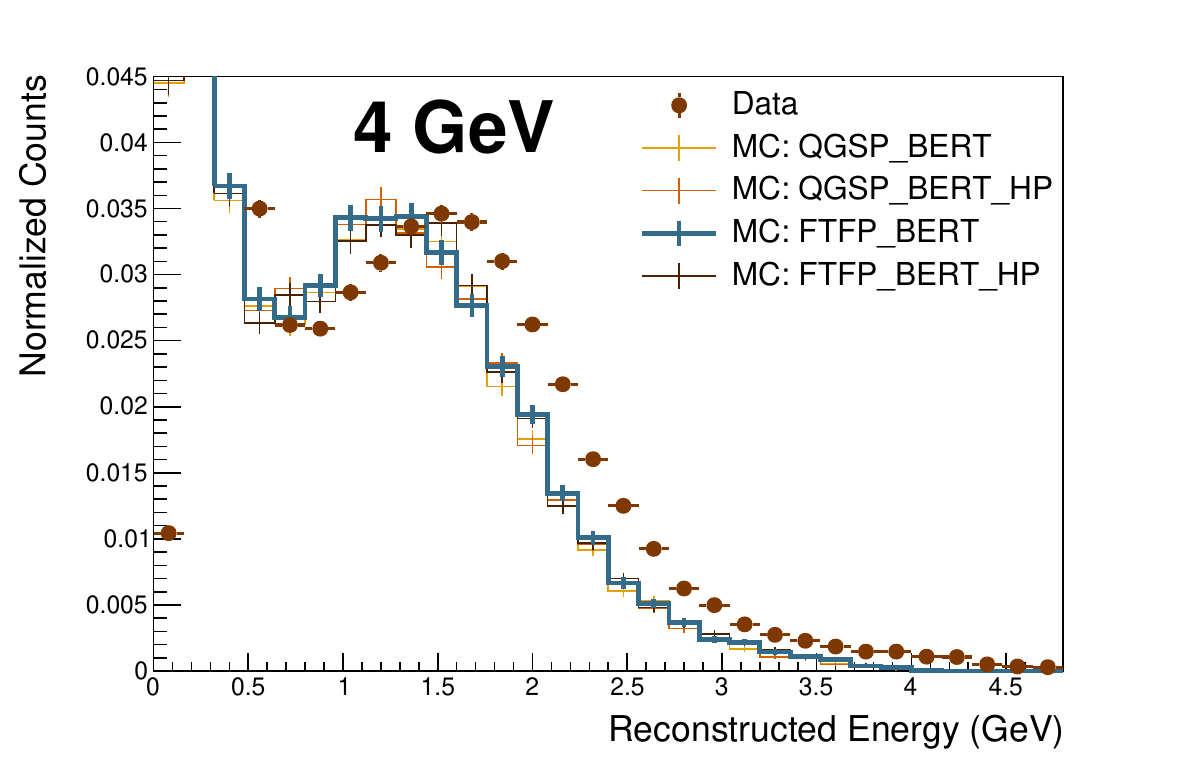}
\qquad
\includegraphics[trim={3 5 55 30},clip,width=.47\textwidth]{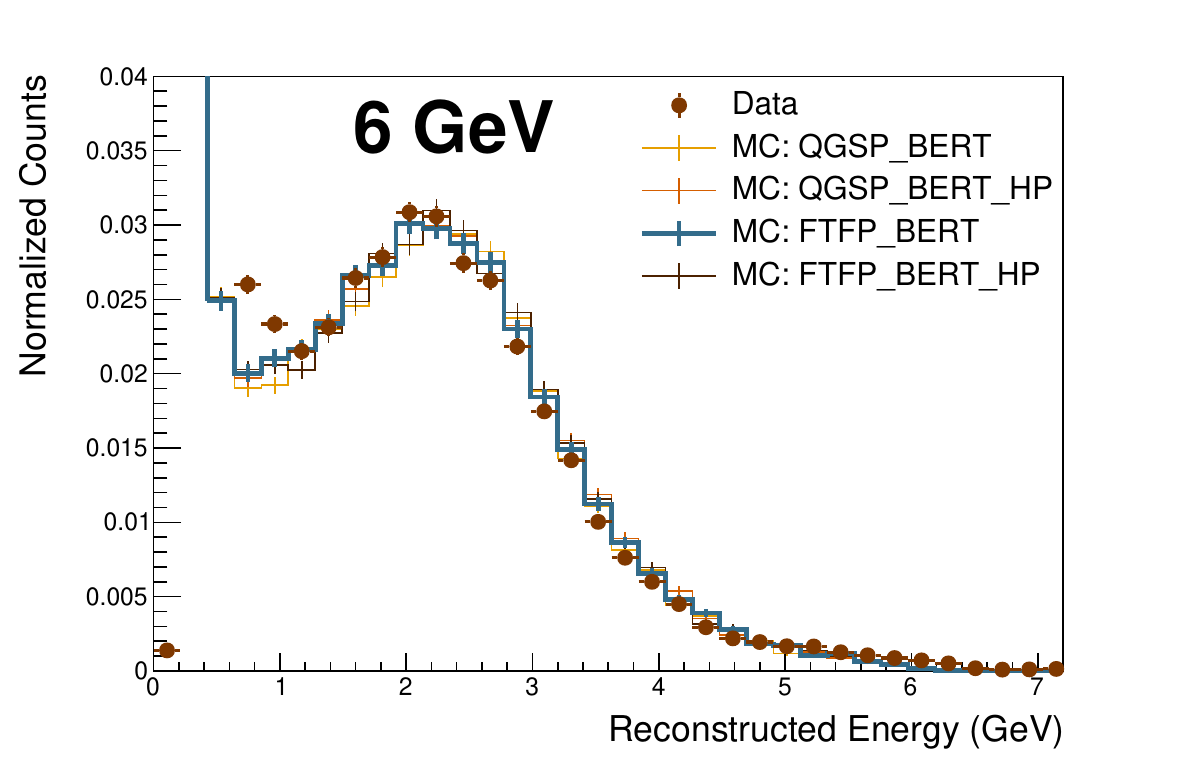}
\qquad
\includegraphics[trim={3 5 55 15},clip,width=.49\textwidth]{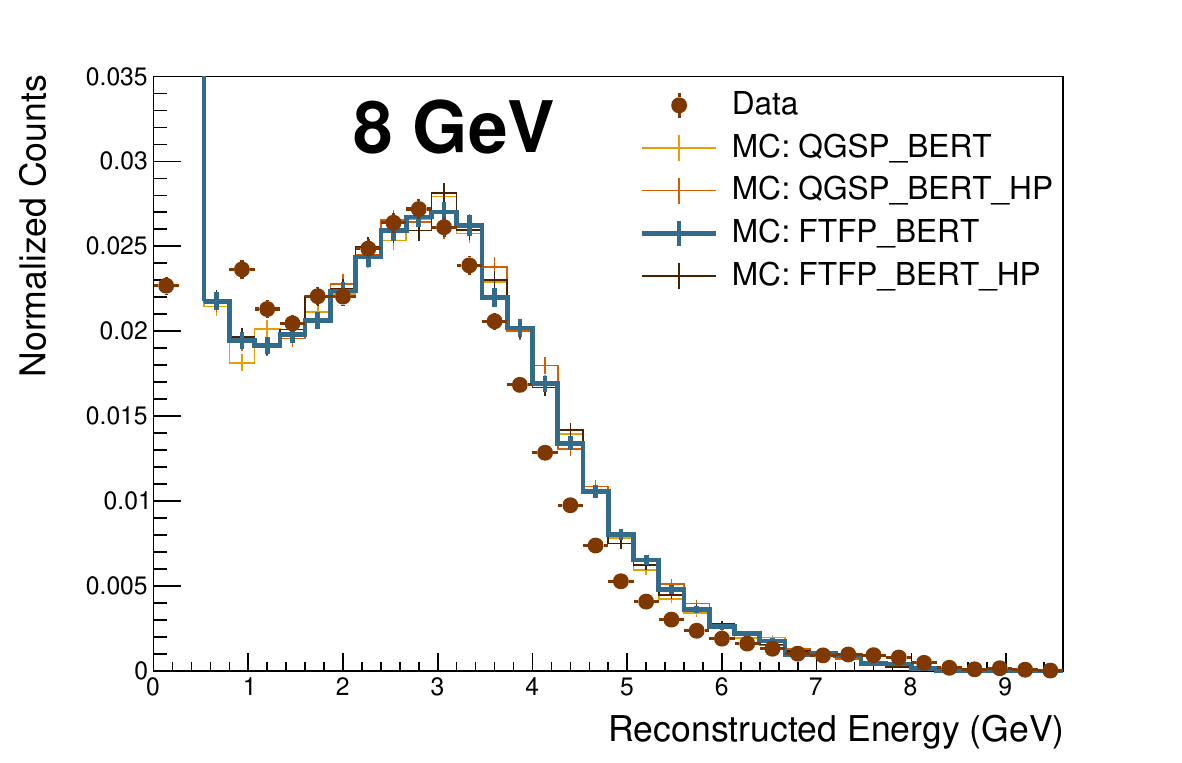}
\caption{Pion energy response for 4\,GeV, 6\,GeV, and 8\,GeV data (circular markers) and simulations (histograms) generated with different GEANT4 physics lists QGSP\_BERT, QGSP\_BERT\_HP, FTFP\_BERT, and FTFP\_BERT\_HP.}
\label{fig:pionSpectrumList}
\end{figure}

\begin{figure}[htbp]
    \centering
    \includegraphics[trim={3 5 55 30},clip,width=.47\textwidth]{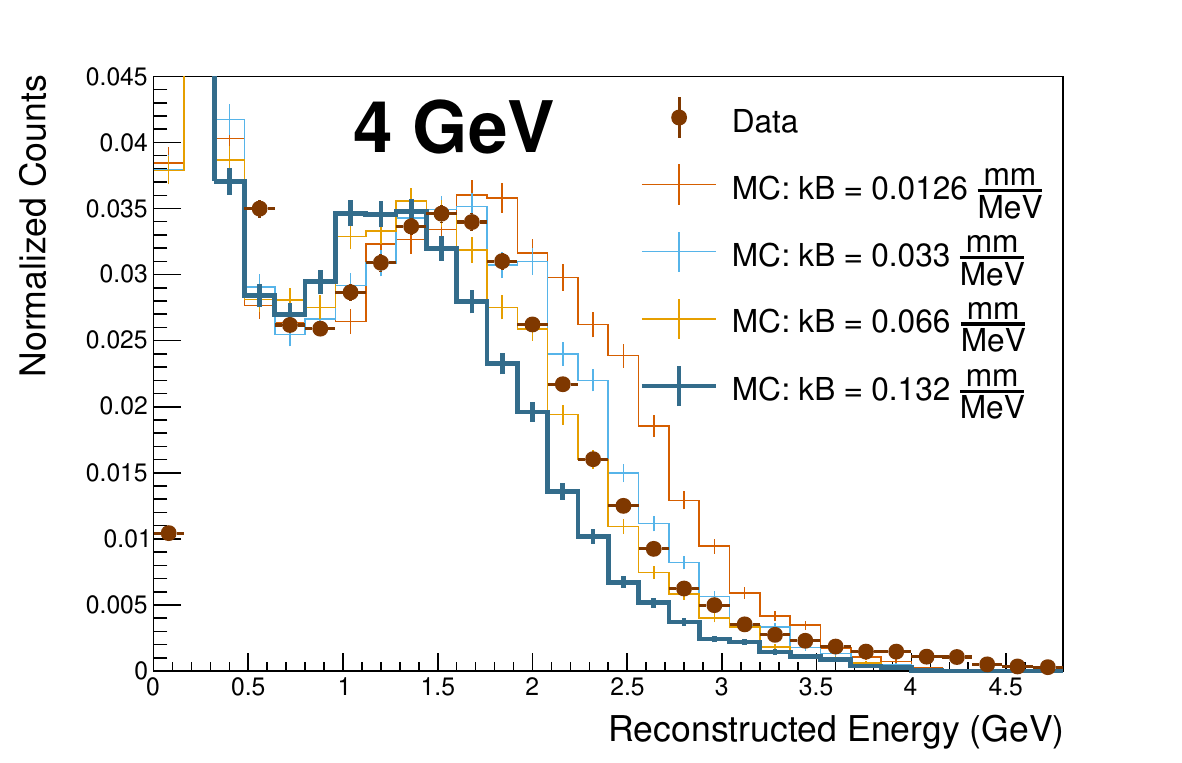}
    \qquad
    \includegraphics[trim={3 5 55 30},clip,width=.47\textwidth]{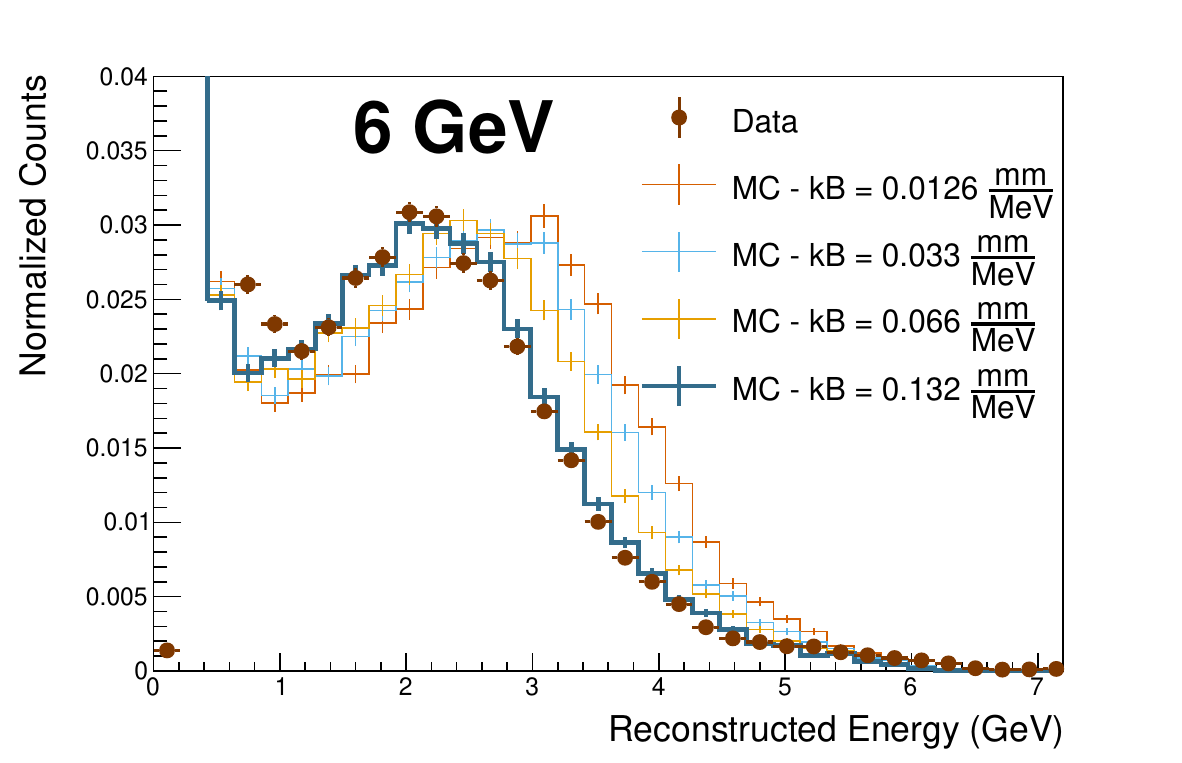}
    \qquad
    \includegraphics[trim={3 5 55 15},clip,width=.49\textwidth]{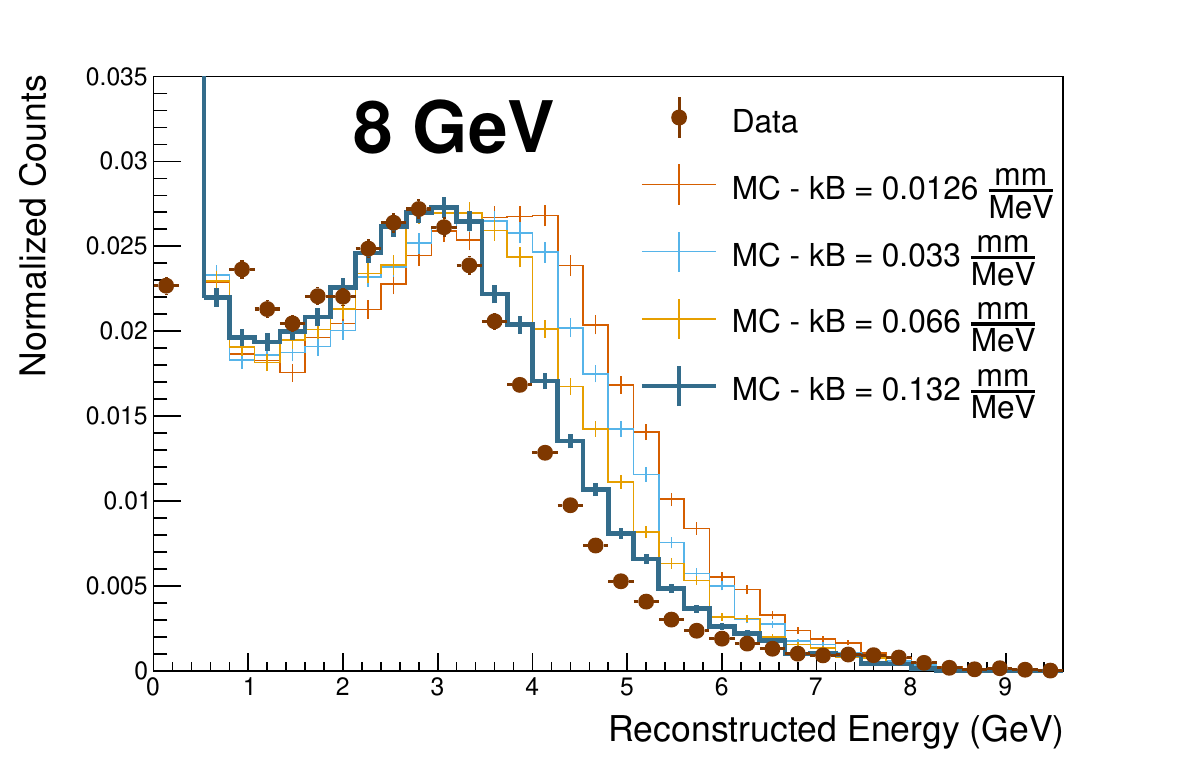}
    \caption{Pion energy response for 4\,GeV, 6\,GeV, and 8\,GeV data (circular markers) and simulations (histograms) with different Birks' constant as marked in the legend.}
    \label{fig:pionSpectrumkB}
\end{figure}

\begin{figure}[htbp]
    \centering
    \includegraphics[trim={3 5 55 30},clip,width=.47\textwidth]{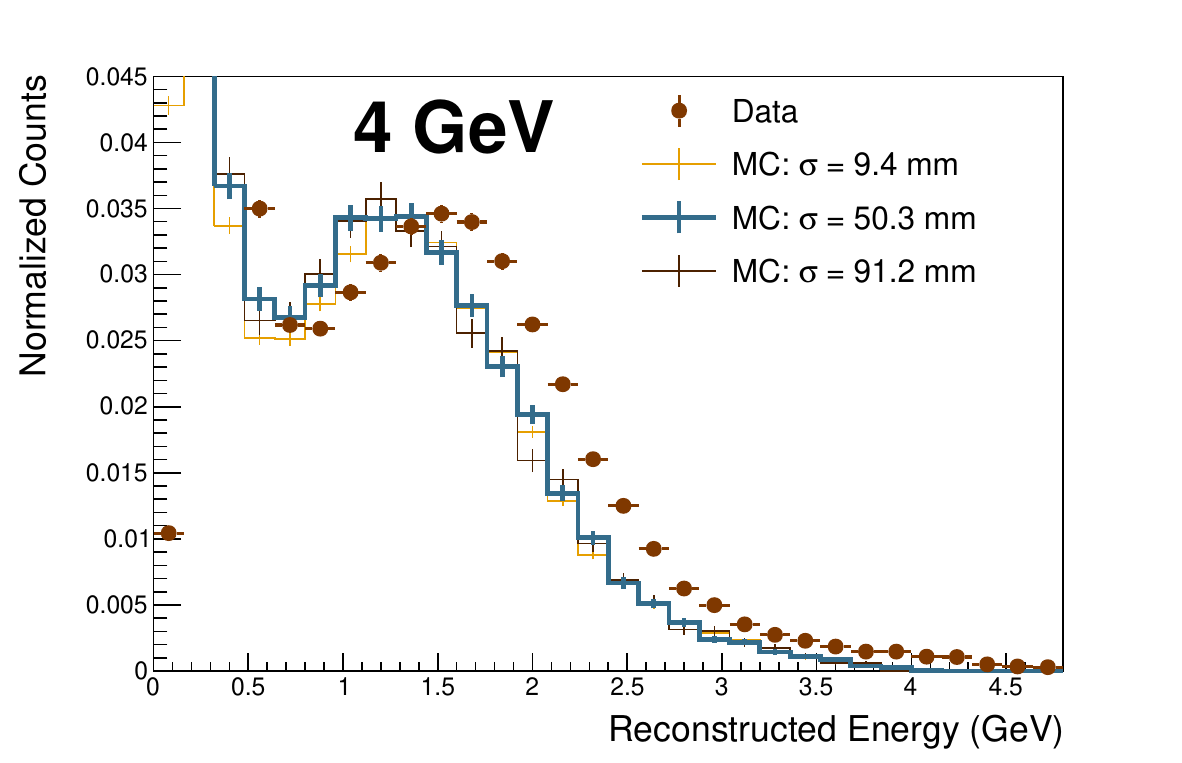}
    \qquad
    \includegraphics[trim={3 5 55 30},clip,width=.47\textwidth]{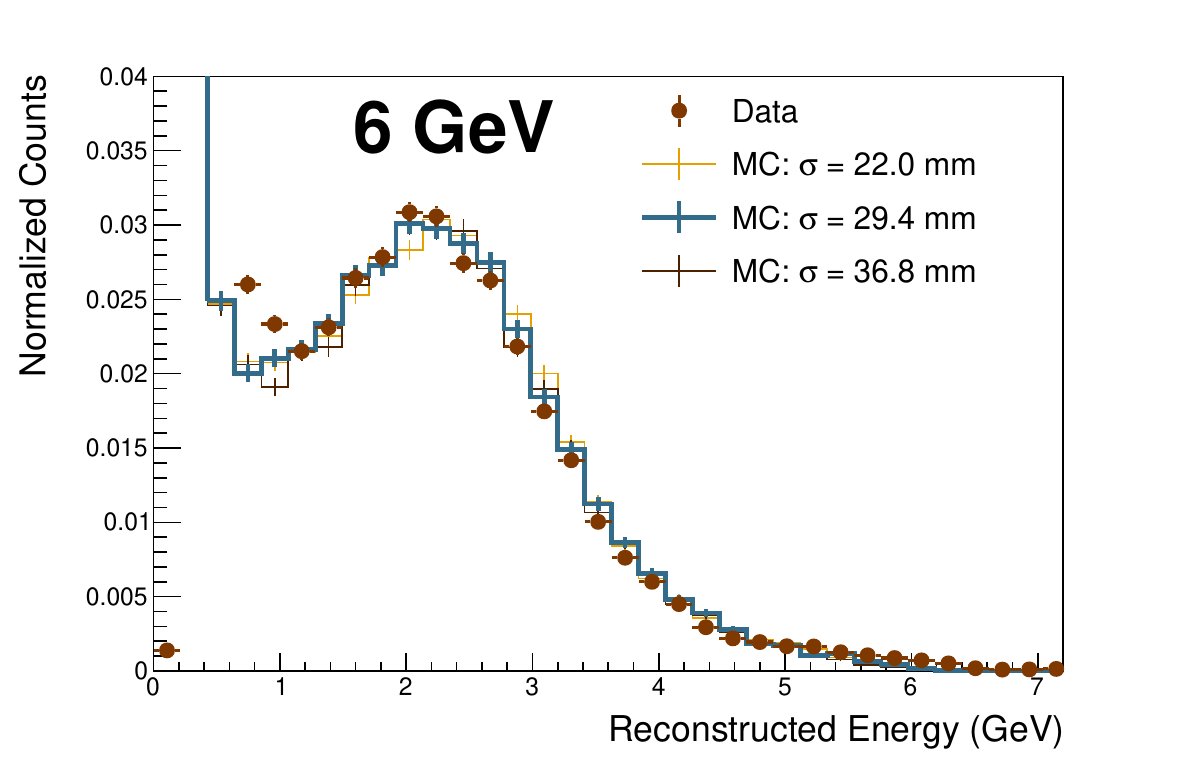}
    \qquad
    \includegraphics[trim={3 5 55 15},clip,width=.49\textwidth]{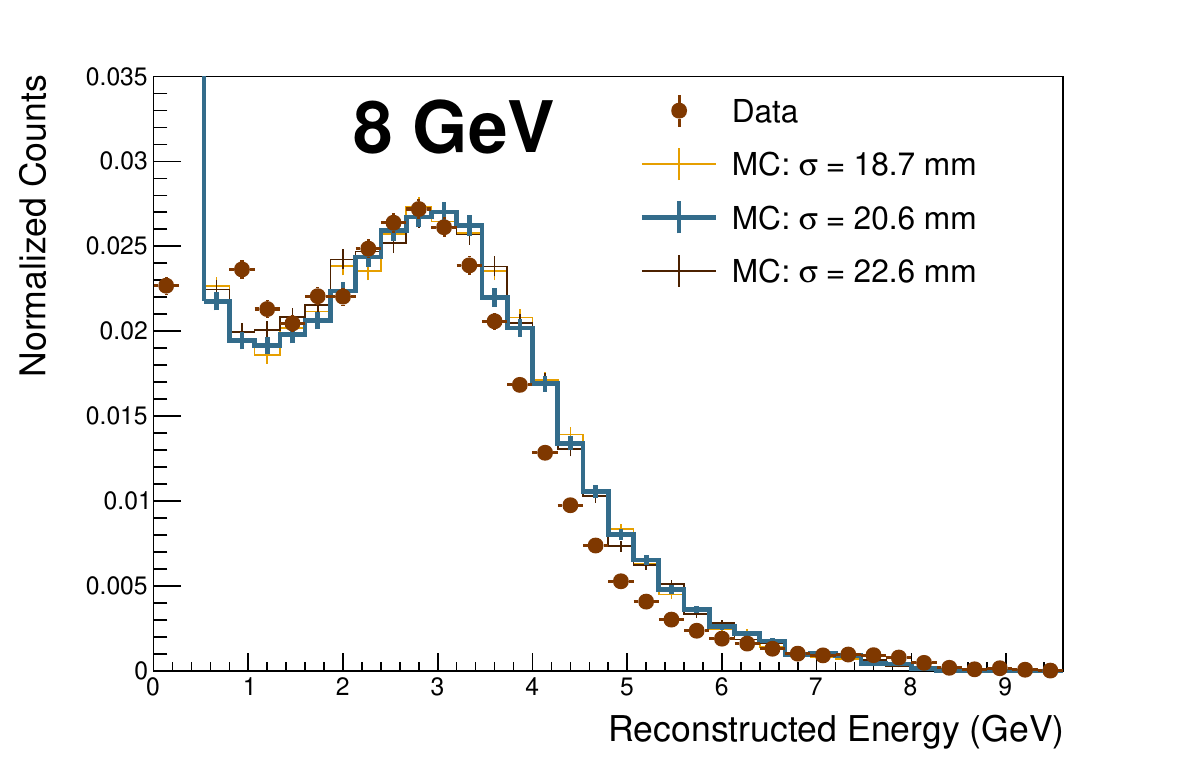}
    \caption{Pion energy response for 4\,GeV, 6\,GeV, and 8\,GeV data (circular markers) and simulations (histograms) with different beam profile spreads as marked in the legend.}
    \label{fig:pionSpectrumProfile}
\end{figure}

This appendix presents figures illustrating the investigation of the pion energy response in simulations described in Sec.~\ref{sec:PionEnergyResponse}.
Figure~\ref{fig:pionSpectrumList} presents the pion energy response for 4\,GeV, 6\,GeV, and 8\,GeV data and simulations with different GEANT4 physics lists. The circular markers represent data for events identified as pions. The energy response has been calibrated using the electron method. The histograms corresponds to simulations generated with the GEANT4 physics lists QGSP\_BERT, QGSP\_BERT\_HP, FTFP\_BERT, and FTFP\_BERT\_HP with nominal beam energy and the ``Adjusted to the data'' beam momentum spread, as shown in Table~\ref{tab:beam_conditions}. The simulated distributions are normalized to the data by scaling them so that the peak height of the energy loss spectrum for showering (non-MIP) pions matches between data and simulation as explained for Fig.~\ref{fig:pionSpectrum}. Figure~\ref{fig:pionSpectrumkB} presents the pion energy response for data and simulations with different Birks' constant values equal to 100\%, 50\%, 25\%, and 9.5\% of its nominal value of 0.132\,mm/MeV. Figure~\ref{fig:pionSpectrumProfile} shows the pion energy response for data and simulations with different beam profile spreads, corresponding to a variation of $\pm1\sigma$ of the extracted beam profiles from the fits presented in Sec.~\ref{sec:FTBF}.

% Bibliography
\clearpage
\bibliographystyle{JHEP}
\bibliography{biblio.bib}
\end{document}